\documentclass[aps,reprint,shortbibliography,nofootinbib,superscriptaddress]{revtex4-1}

\usepackage[utf8]{inputenc}

\usepackage{graphicx}
\usepackage{dcolumn}
\usepackage{color}
\usepackage{amsmath}
\usepackage{enumerate}
\usepackage{cancel}
\usepackage{multirow}
\usepackage{bm}
\usepackage{hyperref}
\usepackage[export]{adjustbox}
\usepackage[nolist,printonlyused]{acronym}

\newcommand{\propdelay}[1]{\mathbf{D}_{#1}}
\newcommand{\tdidelay}[1]{\mathcal{D}_{#1}}
\newcommand{\filter}{\mathcal{F}}

\newcommand{\isig}[1]{\textrm{isi}_{#1}}
\newcommand{\tmig}[1]{\textrm{tmi}_{#1}}
\newcommand{\rfig}[1]{\textrm{rfi}_{#1}}
\newcommand{\isi}{\textrm{isi}}
\newcommand{\tmi}{\textrm{tmi}}
\newcommand{\rfi}{\textrm{rfi}}

\newcommand{\va}[1]{\vec{#1}}
\newcommand{\vu}[1]{\hat{\bf #1}}

\newcommand{\cxx}{C_{XX}(\omega)}
\newcommand{\cxy}{C_{XY}(\omega)}

\definecolor{forestgreen}{rgb}{0.1,0.49,0.07}

\begin{document}

\title{Time-delay interferometry noise transfer functions for LISA}

\author{Dam Quang Nam}
\affiliation{Universit{\'e} Paris Cit\'e, CNRS, Astroparticule et Cosmologie, F-75013 Paris, France}
\author{Yves Lemi\`ere}
\affiliation{Normandie Univ, ENSICAEN, UNICAEN, CNRS/IN2P3, LPC Caen, 14000 Caen, France}
\author{Antoine Petiteau}
\affiliation{IRFU, CEA, Universit\'e Paris-Saclay, F-91191, Gif-sur-Yvette, France}
\affiliation{Universit{\'e} Paris Cit\'e, CNRS, Astroparticule et Cosmologie, F-75013 Paris, France}
\author{Jean-Baptiste Bayle}
\affiliation{University of Glasgow, Glasgow G12 8QQ, United Kingdom}
\author{Olaf Hartwig}
\affiliation{SYRTE, Observatoire de Paris, Universit\'e PSL, CNRS, Sorbonne Universit\'e, LNE, 61 avenue de l’Observatoire, 75014 Paris, France}
\author{Joseph Martino}
\affiliation{Universit{\'e} Paris Cit\'e, CNRS, Astroparticule et Cosmologie, F-75013 Paris, France}
\author{Martin Staab}
\affiliation{Max Planck Institute for Gravitational Physics (Albert Einstein Institute), D-30167 Hannover, Germany}
\affiliation{Leibniz Universit\"{a}t Hannover, D-30167 Hannover, Germany}


\begin{abstract}
The Laser Interferometry Space Antenna (LISA) mission is the future space-based gravitational-wave (GW) observatory of the European Space Agency. It is formed by three spacecraft exchanging laser beams in order to form multiple interferometers. The data streams to be used in order to extract the large number and variety of GW sources are time-delay interferometry (TDI) data. One important processing step to produce these data is the TDI on-ground processing, which recombines multiple interferometric on-board measurements to remove certain noise sources from the data, such as laser frequency noise or spacecraft jitter noise.
The LISA noise budget is therefore expressed at the TDI level in order to account for the different TDI transfer functions applied for each noise source and thus estimate their real weight on mission performance. 
In this study, we present an update model for the beams, measurements and TDI, with several approximations to derive the noise transfer functions. The laser locking and noise correlation are taken into account to see their impact in the transfer functions. 
A methodology for such a derivation has been established in detail, as well as verification procedures against simulated data. It results in a set of transfer functions, which are now used by the LISA project, in particular in its performance model.
Using these transfer functions, realistic noise curves for various instrumental configurations are provided to data analysis algorithms and used for instrument design.
\end{abstract}


\maketitle

\section{Introduction}

The \ac{LISA}~\cite{LISA_Proposal2017, SciRD} is a space-based gravitational wave observatory that aims to measure gravitational waves (GWs) in the millihertz range. The mission is led by \ac{ESA}, with \ac{NASA} as a junior partner, and European member states contributing to both hardware and processing.
\ac{LISA} will enable the observation of parts of the Universe invisible by other means, such as black holes and binaries of compact objects. Furthermore, we will be able to study the very early Universe soon after the big bang, and possibly discover yet completely unknown objects. \ac{LISA} will enhance our knowledge about astrophysics, cosmology and fundamental physics. 
 
\ac{LISA} is composed of three \ac{S/C} in heliocentric orbits, forming an equilateral triangle constellation. 
 The constellation trajectory is in the ecliptic plane at one astronomical unit from the Sun, and leading or trailing Earth on its orbit, with an angular separation of 10 to 30 degrees. This distance from Earth is chosen to minimize \textit{arm-breathing} induced by the Earth while still being close enough to allow communications.
 
The \ac{S/C} exchange laser beams to form multiple interferometers. 
By monitoring the changes in distance between free-falling test-masses inside the spacecraft, \ac{LISA} senses gravitational waves.
 Six laser beams, imprinted by the gravitational waves, connecting the local and distant test-masses, interfere with local laser beams and permit measurements with picometer precision.
Achieving this precision requires the suppression of many technical noise sources, the largest of which is laser frequency noise. It is expected to be several orders of magnitude above \ac{GW} signals.
\ac{TDI} \cite{Estabrook:2000ef, Nayak:2005kb, Vallisneri:2005ji, PhDPetiteau, PhDOtto, PhDBayle, PhDMuratore, PhDHartwig} will suppress this dominant source of noise by 8 orders of magnitude, bringing it below secondary noises and \ac{GW} signals.
The basic idea of \ac{TDI} is to combine time-shifted phase or frequency measurements from the three satellites on-ground to synthesize virtual interferometers which are naturally insensitive to laser frequency noise, but still sensitive to \ac{GW} signals.
Other noise sources that are above the requirements~\cite{SciRD}, need to be suppressed as part of the \ac{TDI} algorithm, such as clock noise. 
Additional algorithms are developed to suppress these noise sources and then integrated to the latest version of the \ac{TDI} algorithm~\cite{Hartwig:2020tdu}. 

To define the instrument performances, two classes of noises are considered: suppressed noises and unsuppressed noises. The suppressed noises are dominant in LISA measurements and should be mitigated by some on-board or offline data processing (e.g., TDI), such as laser frequency noise~\cite{Bayle:2018hnm, Bayle:2021mue}, spacecraft jitter noise, clock and ranging noises~\cite{Hartwig:2020tdu}, tilt-to-length~\cite{Paczkowski:2022nrt}.
The unsuppressed ones are secondary noises, such as test-mass acceleration noise, optical path noises, readout noises, backlink, etc. However, they will constitute the dominant contribution to the LISA instrument noise budget after mitigation of all suppressed noises. This article will present the analytical formulation of how unsuppressed noises propagate through TDI and the validation of these formulations. The analytical transfer function are expressed in TDI second generation, which is required for laser frequency noise suppression in realistic orbits with varying armlengths \cite{Estabrook:2000ef}.

The transfer functions of second generation TDI for the suppressed noises (laser frequency noise \cite{Bayle:2021mue, Bayle:2018hnm}, clock jitter noise \cite{Hartwig:2020tdu, Hartwig:2022yqw} and for the secondary noises \cite{Krolak:2004xp} have already investigated with ideal conditions, i.e., equal armlengths and same statistical property of the same type noises in different \ac{MOSAs}. In this article, we work out the transfer functions of unsuppressed noises without these assumptions. In addition, the impact of laser locking and noise correlation in the transfer function is also investigated. In fact, laser locking and frequency plan were taken into account in some studies of the laser frequency noise reduction, see e.g., \cite{Tinto:2003uk, Valliyakalayil:2021jxd}, but not for the secondary noises. Therefore, we can extend the understanding of the propagation of unsuppressed noise in realistic configuration, which could change the LISA noise budget.

The article is structured as follows. In Sec.~\ref{sec:model}, we introduce the LISA convention, unsuppressed noises, its beam model and measurements, and TDI formulation. The laser locking and some correlation scenarios are also addressed in this section. Then, Sec.~\ref{sec:method} will focus on the methodology to get the power spectral density of the signal as a function of frequency. Some examples show the detailed computation for analytical noise transfer functions with different configurations. At the end of Sec.~\ref{sec:method}, we present the procedure to compare an approximated formulation with instrument simulations performed with \texttt{LISANode}. Section~\ref{sec:results} will be dedicated to the summary of analytical noise transfer functions and the simulation validation. Finally, we conclude in Sec.~\ref{sec:conclusion}.

\section{LISA Model}
\label{sec:model}
\subsection{Convention}
\label{sec:lisa-model:notation}
 
\begin{figure}[h]
 \centering
 \includegraphics[width=0.5\textwidth]{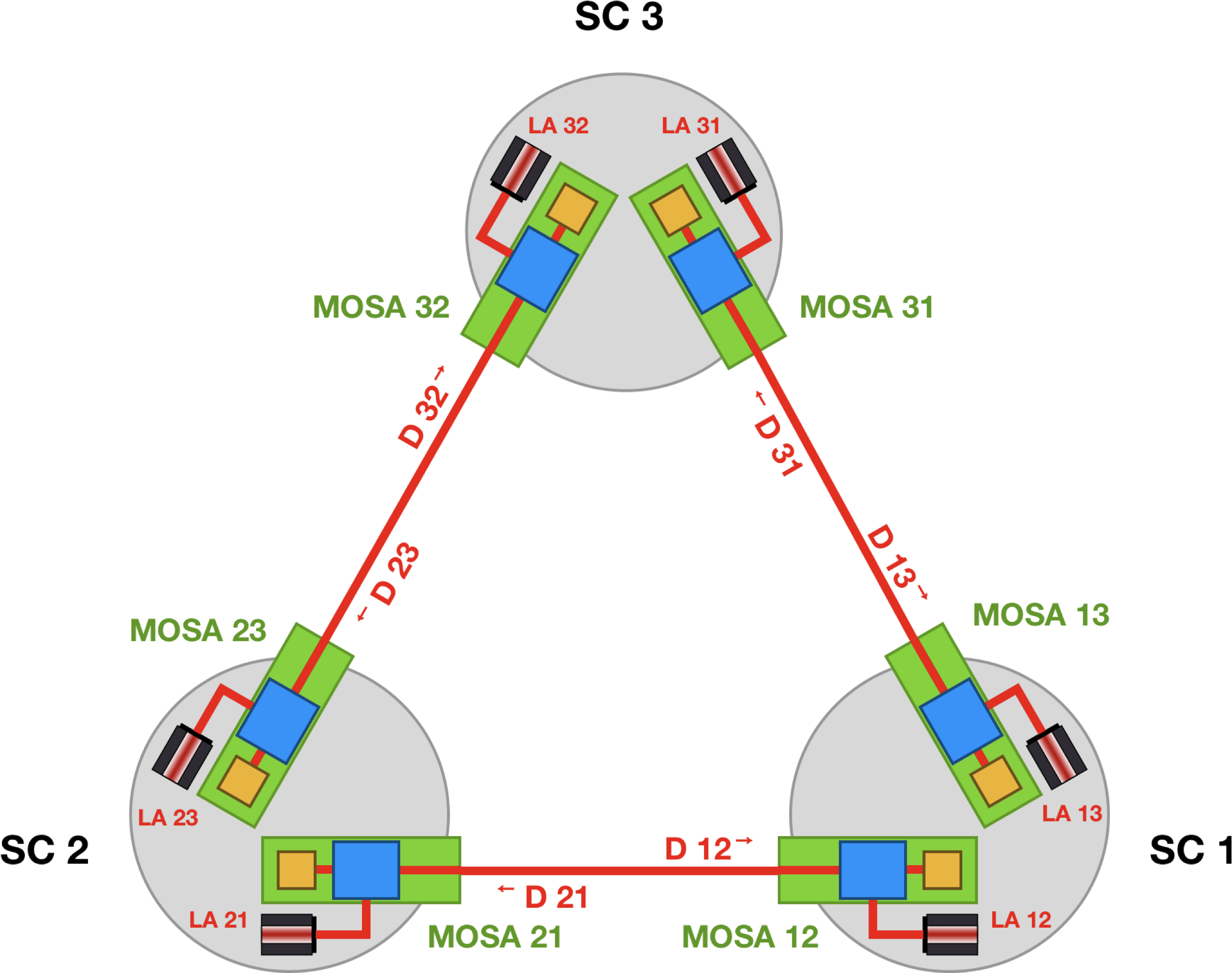}
 \caption{\ac{LISA} constellation convention. The \ac{MOSA} hosted on spacecraft 1 (SC 1) pointing at spacecraft 2 (SC 2) is labeled MOSA$_{12}$.
 Each element hosted on this \ac{MOSA} and the associated laser source will share the same indexes. For example the noise due to the laser associated to the MOSA$_{12}$ will be labeled p$_{12}$.}
 \label{fig:LISA-constellation-convention}
 \end{figure}

 In this article, we follow the convention for the \ac{LISA} constellation proposed by \ac{LISA} Consortium (\cite{Bayle:2022okx}).
 The indexing is summarized on Fig.~\ref{fig:LISA-constellation-convention}. Spacecraft are indexed 1, 2, 3 clockwise when looking down at their solar panels. Each of them hosts two \ac{MOSAs} which include the test-mass and its housing, the optical bench and the telescope. A laser source is associated with each \ac{MOSA}.
 \ac{MOSAs} on each spacecraft are indexed with two numbers $ij$: 
 \begin{itemize}
 \item The first number $i$ is the index of the \ac{S/C} the \ac{MOSA} is mounted on, i.e., the local \ac{S/C}.
 \item The second number $j$ is the index of the \ac{S/C} the \ac{MOSA} points to.
 \end{itemize}
 All subsystems of the \ac{MOSA}, the associated laser and the optical measurements are indexed according to this \ac{MOSA}. There are three main interferometric measurements in each MOSA: \ac{ISI}, \ac{TMI} and \ac{RFI}, which are respectively denoted as isi, tmi, rfi.\footnote{To feed the clock noise reduction algorithm, we also need the sideband measurements in the \isi~and the \rfi~\cite{Hartwig:2020tdu}.} The ISI measurement is monitoring the distance between two optical benches (OBs) in difference spacecraft, while the TMI measures the distance between the test-mass (TM) and OB in the same MOSA. The RFI measurement is used to mitigate the spacecraft jitter noise (relation motion between test-mass and optical bench attached in spacecraft), and to reduce the number of free-running laser as we see later in Sec.~\ref{sec:lisa-model:tdi}. The detailed formulation of these measurements will be presented in Secs.~\ref{sec:lisa-model:beam-model} and~\ref{sec:lisa-model:ifo-measurement}.

 We define $L_{ij}(t)$
 as the light travel time from \ac{S/C} $j$ to \ac{S/C} $i$, in seconds. For the propagation of light, we denote the propagation delay operator\footnote{Technically, since the measurements will be expressed in relative frequency fluctuation units, $\propdelay{ij}$ is a Doppler-delay operator $\propdelay{ij}u(t) = (1 - \dot{L}_{ij}(t)) u(t - L_{ij}(t))$ (see Sec. 7.2 of~\cite{PhDHartwig}).}
 by $\propdelay{ij}$, so that $\propdelay{ij}u(t) = u(t - L_{ij}(t))$ for any time series $u(t)$. 
We also use the \ac{TDI} delay operator $\tdidelay{ij}$, such that $\tdidelay{ij}u(t) = x(t - \hat{L}_{ij}(t))$, where $\hat{L}_{ij}(t)$ is the estimate of the light travel time $L_{ij}(t)$. 
For nested delay operators, we use the short hand notation $d_{i_1 i_2 \dots i_n} \equiv d_{i_1 i_2} d_{i_2 i_3} \dots d_{i_{n-1} i_n}$, where $d$ could be $\propdelay{}$ or $\tdidelay{}$. 
In general, those delay operators are not commutative since light travel times evolve with time. If we use the commutator notation of $[A, B] = AB - BA$ then $[\propdelay{ij}, \propdelay{mn}]u(t) \neq 0$ when $(i,j) \neq (m,n)$. But if delay times or armlengths are assumed to be constant, delay operators become commutative. We will use this to simplify the computation process.

Another process we indicate using an operator is the action of the antialiasing filters, which are used to prevent power folding in the band of interest during decimation.
Its operator is denoted as $\filter$, such as $\filter u(t) = (f * u) (t)$, where the asterisk stands for the convolution of time series $u(t)$ with the filter kernel $f(t)$.
 
 The \ac{GW} signal measured in \ac{ISI}$_{ij}$, caused by the accumulated delay of the beam received on \ac{S/C} $i$ from \ac{S/C} $j$ due to a \ac{GW}, is labeled $H_{ij}$.
 
 The wavelength of laser associated to MOSA$_{ij}$ is $\lambda_{ij}$ and its frequency is denoted as $\nu_{ij} = c / \lambda_{ij}$. We also define the frequency of the laser beam received by MOSA$_{ij}$ from MOSA$_{ji}$ as $\nu_{i \leftarrow j}$. Due to the Doppler shift along the link $L_{ji}$, $\nu_{i \leftarrow j} \neq \nu_{ji}$. 
 The laser frequency $\nu$ is the sum of nominal frequency (carrier or sideband --- THz), an offset frequency (Doppler and laser locking --- MHz) and small fluctuations (noises and GWs --- nHz to Hz).
The interferometric signals in \ac{LISA} are the heterodyne beatnote frequencies, i.e., the frequency differences between the frequencies of associated beams (offsets and small fluctuations). Their signs are (beatnote polarities) $\theta^{\isi}$ and $\theta^{\rfi}$ for \ac{ISI} and \ac{TMI}/\ac{RFI} signals, respectively.
 \begin{equation}
 \left\{
 \begin{array}{ccc}
  \theta^{\isi}_{ij} & = & \textrm{sign}(\omega_{i \leftarrow j} - \omega_{ij}),  \\
  \theta^{\tmi}_{ij} = \theta^{\rfi}_{ij} & = & \textrm{sign}(\omega_{ik} - \omega_{ij}),
 \end{array}
 \right.
 \label{eq:def_theta} 
 \end{equation}
where $\omega = 2 \pi \nu$, $(i, j, k)$ matches every cyclic permutation of $(1, 2, 3)$. In general, $\theta^{\isi}_{ij} \neq - \theta^{\isi}_{ji}$ but $\theta^{\rfi}_{ij} = - \theta^{\rfi}_{ik}$.

 \subsection{List of unsuppressed noises}
 
 The laser frequency noise is the dominant noise source in \ac{LISA}, and reduced by \ac{TDI} postprocessing algorithm (see Sec.~\ref{sec:lisa-model:tdi}). Other noises that are not suppressed by \ac{TDI} or other postprocessing algorithms are called unsuppressed noises. Unsuppressed noises are subdominant (for example with respect to laser frequency noise or clock noise) but once these dominant noises have been suppressed, they contribute to the \ac{LISA} noise budget. It is therefore necessary to study their propagation  through \ac{TDI}. 
 
 The measurements will be either in phase or frequency, or a mixture of both. The final choice is not yet made. Since the noises we are interested are expressed as small fluctuations (phase or frequency), we will assume that the measurements are in relative frequency fluctuations. It is also the unit used for most of the \ac{GW} analyses.

 We will denote the \ac{LISA} instrumental noises as follows:
 \begin{itemize}
  \item $p_{ij}$: laser frequency noise (free-running or locked, see Sec.~\ref{sec:lisa-model:freq-plan}) of the laser on \ac{MOSA} ${ij}$;
  \item $\delta_{ij} = \va{\delta}_{ij} . \vu{n}_{ji} / c$  : projection of test-mass $ij$ jitter noise vector $\va{\delta}_{ij}$ onto the sensitive axis. $\vu{n}_{ji}$ is the reference axis for the \ac{MOSA} $ij$, i.e.,  from test-mass to \ac{OB} (see Figs~\ref{fig:LISA-constellation-convention} and~\ref{fig:convention_directions_beam_motion}). 
We assume that all measurements are in fractional frequency units. The test-mass jitter noise is expressed in velocity (m/s), so we need the factor $1/c$ (see~\cite{PhDBayle} for the detailed derivation);
  \begin{figure}[h]
 \centering
 \includegraphics[width=0.5\textwidth]{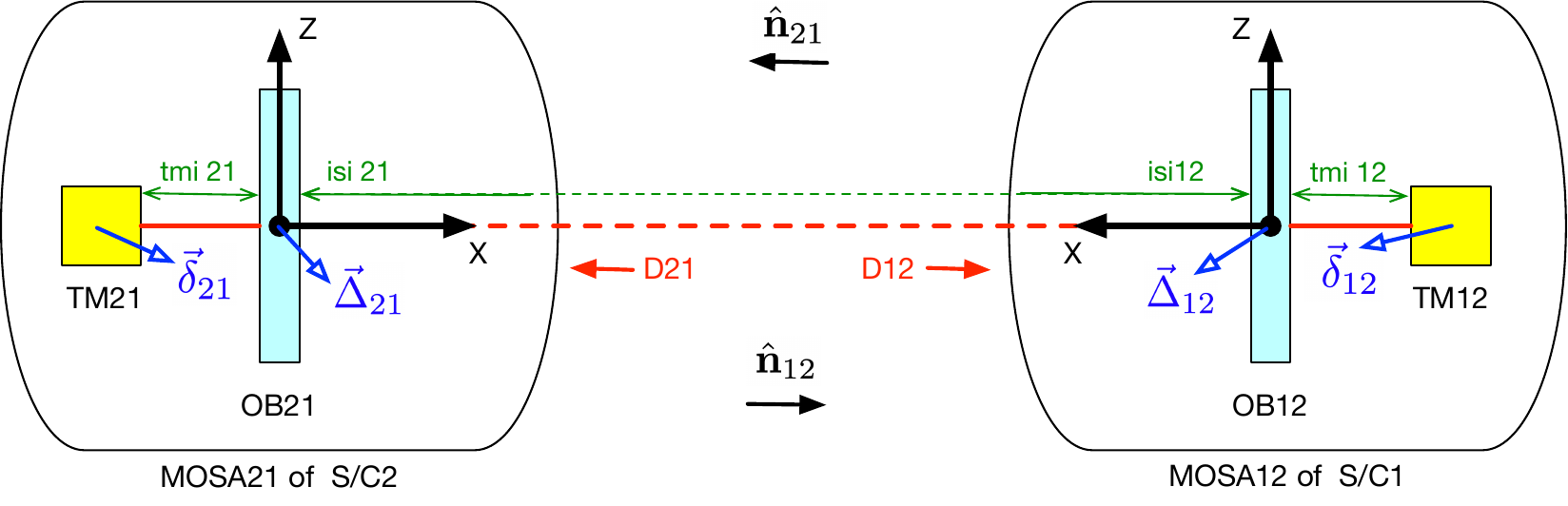}
 \caption{Conventions for direction of beams and motions for \ac{MOSA} $12$ and \ac{MOSA} $21$. The reference \textsf{X} axis for \ac{MOSA} 12 is equal to $\vu{n}_{21}$. }
 \label{fig:convention_directions_beam_motion}
 \end{figure}
  \item $\Delta_{ij} = \va{\Delta}_{ij} . \vu{n}_{ji} / c$: projection of \ac{MOSA} ${ij}$ jitter noise vector $\va{\Delta}_{ij}$ onto the sensitive axis (longitudinal axis);
  \item $ \mu^{x}_{ij \rightarrow ik} $: backlink noise for the $x$ measurement,  $x \in \{\tmi , \rfi \}$. This noise is dominated by straylight in the optical fiber connecting two \ac{MOSAs} of the same \ac{S/C} [from \ac{OB}$_{ij}$ to \ac{OB}$_{ik}$, where $(i, j, k)$ is the set of combination of $(1, 2, 3)$]. In general, this noise is nonreciprocal, i.e., $ \mu^{x}_{ik \rightarrow ij} \neq  \mu^{x}_{ij \rightarrow ik}$.
  \item $ N^{ro}_{x,ij} $: readout noise for the $x$ measurement of \ac{OB}$_{ij}$, $x \in \{ \isi , \tmi , \rfi \}$ ;
  \item $N^{\textrm{op}}_{\alpha,ij}$ : generic \ac{OP} noise term due to optical path fluctuations on \ac{OB} ${ij}$. $\alpha$ refers to:
    \begin{itemize}
    \item $TX/\isi$: \ac{OP} noise on the beam transmitted to the distant \ac{S/C} induced by the sending \ac{S/C};
    \item $RX/\isi$: \ac{OP} noise on the beam received from the distant \ac{S/C} induced by the receiving \ac{S/C};
    \item $\tmi$: \ac{OP} noise on adjacent beam in the \ac{TMI} measurement;
    \item $\rfi$: \ac{OP} noise on adjacent beam in the \ac{RFI} measurement;
    \item $loc/\isi$:  \ac{OP} noise on local beam in the \ac{ISI} measurement;
    \item $loc/\tmi$: \ac{OP} noise on local beam in the \ac{TMI} measurement;
    \item $loc/\rfi$: \ac{OP} noise on local beam in the \ac{RFI} measurement.
    \end{itemize}
 \end{itemize}

 \subsection{Beam modeling}
 \label{sec:lisa-model:beam-model}
 In order to model the interferometric measurement, we start by modeling the beams that interfere in terms of combination of noises.
 The main six beams of the three interferometers in the \ac{MOSA} 12 are described as

 \begin{subequations}
 \begin{align}
 b_{\isi ,21 \rightarrow 12} & =  \propdelay{12}\left[ p_{21} + N^{\textrm{op}}_{TX/\isi,21} - \frac{1}{c} \vu{n}_{12}.\va{\Delta}_{21} \right] \nonumber \\
  & \quad + H_{12}
 - \frac{1}{c} \vu{n}_{21}.\va{\Delta}_{12} +  N^{\textrm{op}}_{RX/\isi,12} \\
 b_{\tmi, 13\rightarrow 12} & =   p_{13} + \mu^{\tmi}_{13 \rightarrow 12} 
 + N^{\textrm{op}}_{\tmi ,12}\\
 b_{\rfi, 13\rightarrow 12} & =  p_{13} + \mu^{\rfi }_{13 \rightarrow 12} +  N^{\textrm{op}}_{\rfi ,12}\\
 b_{\isi, 12\rightarrow 12} & =  p_{12} +  N^{\textrm{op}}_{loc/\isi , 12}\\
 b_{\tmi, 12\rightarrow 12} & =  p_{12} 
    + \frac{2}{c} \vu{n}_{21} .(\va{\Delta}_{12} - \va{\delta}_{12})
    + N^{\textrm{op}}_{loc/\tmi , 12} \label{eq:beam_tmi_12_12}\\
 b_{\rfi , 12\rightarrow 12} & =  p_{12} +  N^{\textrm{op}}_{loc/\rfi ,12}
 \end{align}
 \end{subequations}
 where 
 \begin{itemize}
 \item $b_{\isi ,21 \rightarrow 12}$ is the beam from \ac{MOSA} $21$ received by \ac{MOSA} $12$,
 \item $b_{\rfi ,13 \rightarrow 12}$ and $b_{\tmi ,13 \rightarrow 12}$ are the beams propagating from \ac{MOSA} $13$ to \ac{MOSA} $12$ through the backlink, which respectively contribute to \ac{RFI} and \ac{TMI} measurements.
 \item $ b_{x, 12\rightarrow 12}$ are the local beams of the \ac{MOSA} $12$ with $x \in \{\isi , \tmi , \rfi \}$.
 \end{itemize}
 
 In the current design, the local beam of the \ac{TMI}, $b_{\tmi, 12\rightarrow 12}$, is bouncing on the test-mass. 
 The sign convention is such that if the test-mass moves toward the \ac{OB}, i.e., $\va{\delta}_{12}$ points in the positive direction (\textsf{X} axis of \ac{MOSA} 21, $\vu{n}_{21}$, see Fig.~\ref{fig:convention_directions_beam_motion}), the optical path on the beam $b_{\tmi, 12\rightarrow 12}$ decreases.
 If the \ac{OB} moves away from the test-mass, i.e., $\va{\Delta}_{12}$ points in the positive direction, the optical path on the beam $b_{\tmi, 12\rightarrow 12}$ increases while it decreases on $b_{\isi, 21\rightarrow 12}$.

 The beams in \ac{MOSA} 13 are constructed in the same way. One can easily write them from the formulas of \ac{MOSA} 12 by replacing index 2 by 3 everywhere.
 The beams in the other \ac{MOSAs} can be deduced by circular permutation of indices $(1 \rightarrow 2 \rightarrow 3 \rightarrow 1)$.

 \subsection{Interferometer measurement}
 \label{sec:lisa-model:ifo-measurement}
 
 Using those beams, we can construct the three main interferometric measurements, for example in the \ac{MOSA} 12, as follows
 \begin{equation}
 \left\{
 \begin{array}{ccc}
  \isig{12} & = & \filter \left[ \theta_{12}^{\isi}  \left( b_{\isi ,21 \rightarrow 12} - b_{\isi ,12 \rightarrow 12} \right) + N^{ro}_{\isi, 12} \right]  \\[6pt]
  \tmig{12} & = & \filter \left[ \theta_{12}^{\rfi}  \left( b_{\tmi , 13 \rightarrow 12}  - b_{\tmi , 12 \rightarrow 12}  \right) + N^{ro}_{\tmi ,12} \right] \\[6pt]
  \rfig{12} & = & \filter \left[ \theta_{12}^{\rfi}  \left( b_{\rfi , 13 \rightarrow 12} - b_{\rfi , 12 \rightarrow 12} \right) + N^{ro}_{\rfi ,12} \right].
 \end{array}
  \right.
 \end{equation}

As indicated before, we are interested in the small fluctuations from noises and GWs, and so neglect the beatnote offsets in the intereferometric measurements\footnote{The beatnote polarities $\theta^{\isi}, \theta^{\rfi}$ depend on the total laser frequencies of the interefering beams. However, they only define the signs of the beatnote measurements.}. All measurements are expressed in relative frequency fluctuation units.
In phase units, these equations are similar, with additional conversion factors.

\subsection{Correlations}\label{sec:lisa-model:corr-scen}

 Even though, the impact of correlations has been discussed in early TDI studies~\cite{Sylvestre:2003in}. In most studies, as for example~\cite{Robson:2018ifk,Larson:1999we,LISA_Proposal2017,SciRD}, the LISA Instrument noise performance are assessed as uncorrelated single link contribution from optical measurement system and test-mass acceleration.
 This assumption simplifies the calculation of noise propagation but may induce non-negligible errors in the estimation of \ac{LISA} performances. 
To quantitatively estimate the deviation from the ideal case, we will consider some generic scenarios of correlation in this study. Furthermore, we can split the noises into two parts, the correlated and uncorrelated terms, and derive their transfer functions separately. 
 
 One obvious correlation scenario is related to the thermo-mechanical optical path noises in the telescope\footnote{While the optical path noise enters in the \ac{ISI} measurements in the same way as the \ac{MOSA} jitter noise, it is not canceled by the time-delay interferometry algorithm, as described later in Sec.~\ref{sec:lisa-model:tdi}, because it does not appear in the \ac{TMI} measurements.}. 
 Since the same telescope is used for both sending and receiving beams, it will imprint an identical noise at the \ac{ISI} beam, located at both end of a link. 
 The optical path noise on the emitted beam $N^{OP}_{TX/\isi,ij}$ and the received beam $N^{OP}_{RX/\isi,ij}$ in the telescope of \ac{MOSA} $ij$ are fully correlated:
 \begin{equation}
 N^{OP}_{TX/\isi,ij} = N^{OP}_{RX/\isi,ij}.
 \label{eq:tel_fully_correlation_scen}
 \end{equation}
  
 Another possible correlation scenario is related to test-mass acceleration noise. The two test-masses share the same \ac{S/C} and thus will likely have correlated source of noises like temperature driven noises (stiffness, symmetric outgassing), cross-talk of \ac{S/C} jitter, coupling with local and interplanetary magnetic fields or local gravity field fluctuation. 
 We express it by the following correlation relation
 \begin{equation}
\va{\delta}_{ij}.\vu{n}_{ji} = \gamma \, \va{\delta}_{ik}.\vu{n}_{ki},
 \label{eq:tm_corr_scen}
 \end{equation}
 where $\gamma$ is the correlation factor and $(i,j,k)$ can be any permutation of $(1,2,3)$. $\gamma$ is 1 in the case of fully correlated noise, or -1 in case of anticorrelation. We will derive the propagation of the fully correlated acceleration noise in Sec.~\ref{sec:results}.
 The transfer functions for fully correlated and anticorrelated acceleration noise, fully correlated and anticorrelated adjacent (same \ac{S/C}) interferometer noise and fully correlated optical path noise at the same telescope are also given (see Tables~\ref{tab:summary_unsup_table_XYZ} and~\ref{tab:summary_unsup_table_AET}).

 \subsection{Frequency planning --- Laser locking scheme}
 \label{sec:lisa-model:freq-plan}
 
 The intersatellite separation distance varies in time due to orbital dynamics. As a consequence, the laser beam coming from the distant \ac{S/C} is frequency-shifted by about 10 MHz according to the Doppler effect.
The laser frequencies used for the interferometric measurement are slightly offset. 
There is a time evolution of the beatnote between the two beams used to measure phase shift via heterodyne interferometry.
 
The optical measurement system tracks the beatnote frequencies in the range of 5 to 25 MHz, which is not compatible with free running lasers and Doppler-shifted beams.
To accomodate this constraint, we lock the lasers by controlling the frequency of a laser (therefore the beatnote frequencies) such that they remain equal to a preprogrammed reference value~\cite{FreqPlanGH}. We use the \ac{RFI} measurement to phase-lock a laser with its adjacent laser in the same \ac{S/C} (local locking), and the \ac{ISI} signal to lock the local laser to the distant laser (distant locking). In the end, 5 of 6 lasers will be locked on the primary laser. 
In this study, we assume that laser frequency control works perfectly so the locking beatnote offset, laser frequency offset plus the Doppler shift if it is distant locking, is exactly equal to the desired value. 
We also do not consider the beatnote offset in the interferometric measurement, as discussed in Sec.~\ref{sec:lisa-model:ifo-measurement}.
The constraint equation of the beatnote fluctuation is used without filter since the laser locking control loop operates at high frequency before measurements are filtered and downsampled~\cite{PhDHartwig}. 

\begin{figure}[h]
 \centering
 \includegraphics[width=0.43\textwidth]{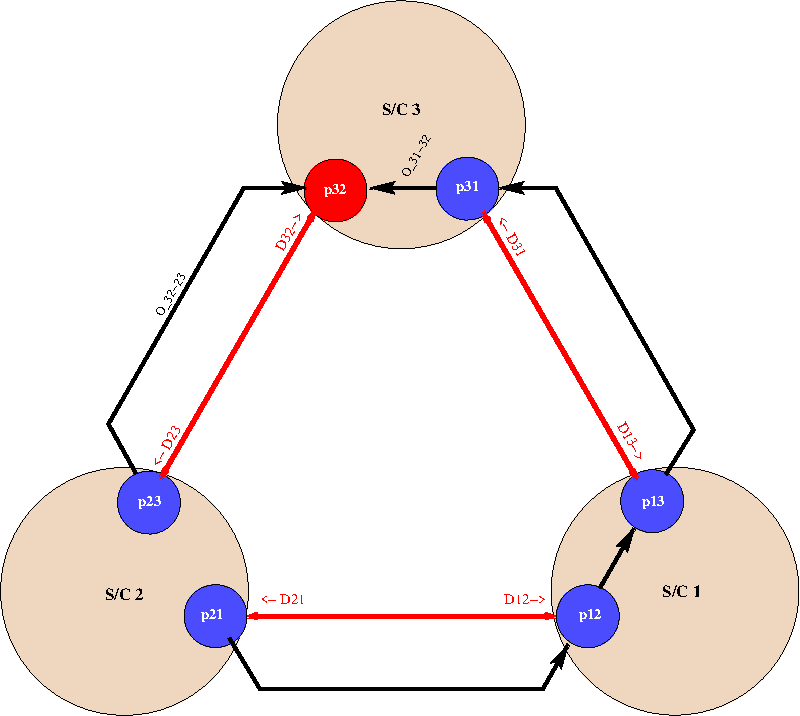}
 \caption{Frequency planning configuration N4-32 (cfg\_N2c in~\cite{FreqPlanGH}). The primary laser is 32 with frequency fluctuations $p_{32}$. The other lasers are locked via \ac{RFI} measurements (31 and 12) or via \ac{ISI} measurements (13, 21 and 23).}
 \label{fig:LISA-planning}
 \end{figure}

 In this study, the configuration N4-32 (cfg\_N2c in~\cite{FreqPlanGH}) has been used\footnote{We used N4-32 because it was the preferred configuration when this study started. Currently the preferred configuration is N1-12 but this does not change the final results which are independent of the locking configuration.}. 
 The detailed phase-locking is shown on Fig.~\ref{fig:LISA-planning}. The constraints on the beatnote fluctuations are
 \begin{subequations}
 \begin{align}
  \isig{21}^{\xcancel{\filter}}    & =  0,\\
  \rfig{31}^{\xcancel{\filter}} & =  0,\\
  \isig{13}^{\xcancel{\filter}}   & =  0,\\
  \rfig{12}^{\xcancel{\filter}} & =  0,\\
  \isig{23}^{\xcancel{\filter}}   & =  0,
 \end{align}
 \end{subequations}
where $\xcancel{\filter}$ indicates the unfiltered and nondownsampled signal. These equations yield the following formulation for the five locked laser frequency fluctuations:

\begin{subequations}
\begin{align}
  p_{23}    & = \theta_{23}^{\isi} N^{ro}_{\isi,23}  + b_{\isi, 32 \rightarrow 23} - N^{op}_{loc/\isi,23} ,\\
  p_{31}    & = \theta_{31}^{\rfi} N^{ro}_{\rfi, 31}+ b_{\rfi, 32 \rightarrow 31} - N^{op}_{loc/\rfi,31} ,\\
  p_{13}    & = \theta_{13}^{\isi} N^{ro}_{\isi,13} + b_{\isi, 31 \rightarrow 13} - N^{op}_{loc/\isi,13} , \\
  p_{12}    & = \theta_{12}^{\rfi} N^{ro}_{\rfi,12} + b_{\rfi, 13 \rightarrow 12} - N^{op}_{loc/\rfi,12} ,\\
  p_{21}    & =  \theta_{21}^{\isi}  N^{ro}_{\isi,21} + b_{\isi, 12 \rightarrow 21} -  N^{op}_{loc/\isi,23} .
\end{align}
\label{eq:freq_plan_N2c}
\end{subequations}

 \subsection{Time-delay interferometry formulation}
 \label{sec:lisa-model:tdi}
 Due to the unequal armlengths entering the \ac{ISI} measurements, laser noise cannot be canceled out when two beams interfere at the photodiode. 
 While the lasers are prestabilized, the laser frequency noise is still too high and will be a few orders of magnitude above the mission requirements~\cite{MRD}. The postprocessing algorithm called \ac{TDI} will reduce the laser frequency noise below the requirements by building an equal-arm interferometer (combining time-shifted raw phase measurements). It has been showed that \ac{TDI} preserves the gravitational wave signal~\cite{Petiteau:2006me,Vallisneri:2005ji,Tinto:2014lxa}.

 The \ac{TDI} formulation involves several steps, which yield the \ac{TDI} variables. The first step is to suppress the spacecraft motion (also dubbed optical bench displacement) noise $\Delta_{ij}$, by introducing the intermediary variables $\xi_{ij}$. The idea is to extract the spacecraft jitter noise from the test-mass interferometer (TMI) measurement by subtracting it with reference (RFI) one, so that the laser frequency noise both contained in \ac{TMI} and \ac{RFI} are canceled out as the result. Then we subtract that result by the \ac{ISI} measurement, which also contains the same noise term, to annihilate the spacecraft jitter noise.
 \begin{eqnarray}
  \xi_{12} & = & \isig{12} - \theta_{12}^{\isi} \theta_{12}^{\rfi} \frac{\lambda_{12}}{\lambda_{21}} \frac{\tmig{12}(t) - \rfig{12}(t)}{2} \nonumber \\
  & & \qquad - \theta_{12}^{\isi} \theta_{21}^{\rfi} \frac{\tdidelay{12} \left[ \tmig{21}(t) - \rfig{21}(t) \right] }{2}, \label{eq:tdi_xi12}\\
  \xi_{13} & = & \isig{13} - \theta_{13}^{\isi} \theta_{13}^{\rfi} \frac{\lambda_{13}}{\lambda_{31}} \frac{\tmig{13}(t) - \rfig{13}(t)}{2} \nonumber \\
  & & \qquad - \theta_{13}^{\isi} \theta_{31}^{\rfi} \frac{\tdidelay{13}\left[ \tmig{31}(t) - \rfig{31}(t) \right]}{2}. \label{eq:tdi_xi13}
 \end{eqnarray}
 Then, we can build the second intermediary variables to reduce the number of laser noises by half using the \ac{RFI} measurements.
 \begin{eqnarray}
  \eta_{12}(t) & = & \theta_{12}^{\isi} \xi_{12}(t) +  \frac{\tdidelay{12} \left[ \theta_{21}^{\rfi} \rfig{21}(t) - \theta_{23}^{\rfi} \rfig{23}(t) \right]}{2}, \label{eq:tdi_eta12} \\
  \eta_{13}(t) & = & \theta_{13}^{\isi} \xi_{13}(t) -  \frac{\theta_{13}^{\rfi} \rfig{13}(t) -  \theta_{12}^{\rfi} \rfig{12}(t)}{2}. \label{eq:tdi_eta13}
 \end{eqnarray}
 
 From the intermediary variables $\eta_{ij}$, we can build the TDI variables that reduce laser noise.
 Several TDI combinations exist~\cite{Vallisneri:2005ji,Muratore:2021uqj,PhDHartwig,PhDMuratore}. 
In this article, we focus on the second generation Michelson variables $X_2, Y_2, Z_2$, where each of the two virtual beams of the \ac{TDI} Michelson~\cite{Vallisneri:2005ji}, visits both distant spacecraft twice. We compute $X_2$ as 
 \begin{eqnarray}
 X_{2} & = & 
        \left(1 - \tdidelay{12131}\right)
        \left[\left(
        \eta_{13} + \tdidelay{13} \eta_{31} \right)+
        \tdidelay{131}\left(\eta_{12} + \tdidelay{12} 
        \eta_{21} \right)\right]\nonumber \\          
        & & - \left(1 - \tdidelay{13121}\right)
        \left[\left(
        \eta_{12} + \tdidelay{12} \eta_{21} \right)+
        \tdidelay{121}\left(\eta_{13} + \tdidelay{13} 
        \eta_{31} \right)\right]. \nonumber \\ \label{eq:TDIX20}  
 \end{eqnarray}
The other two Michelson combinations $Y_2$ and $Z_2$ are derived from this equation by circularly permuting all indices.

 \section{Methodology}
 \label{sec:method}
 In this section, we introduce our method to compute the \ac{TDI} transfer function of the noise propagation, using as an example test-mass acceleration noise. 
 Approximations for the simplified result are then justified. Finally, we validate the analytic transfer functions of several noises using the \texttt{LISANode} simulator.
 
 \subsection{PSD/CSD computation}\label{sec:method:PsdCsdComputation}
 We will briefly introduce a method for calculating the spectral density, which follows the procedure used in the software~\cite{psdpipweb}. 
 The \ac{CSD} of two signals $u(t)$ and $v(t)$ can be defined as 
 \begin{equation}
 S_{uv}(f) = CSD[u,v] = \lim\limits_{T \rightarrow \infty} \dfrac{1}{T} \tilde{u}^*_T(f) \tilde{v}_T(f)
 \equiv \langle \tilde{u}^*(f) \tilde{v}(f) \rangle.
 \end{equation} 
 where $\tilde{u}(f)$ is the Fourier transform of $u(t)$ at the frequency $f$, $a^*$ denotes the complex conjugate of any function $a$. $u_T(t)$ is $u(t)$ restricted to a time window of duration $T$. $\tilde{u}_T(f)$ is the Fourier transform of $u_T(t)$. 
 
It is obvious to show that $S_{vu}(f)$ is just the complex conjugate of $S_{uv}(f)$.  The \ac{PSD} of some stationary signal $u(t)$ is $S_{uu}$. It describes the energy contained in the signal $u(t)$ around the frequency $f$. Details on the calculation of the \ac{PSD} and its statistic properties  are provided in appendix~\ref{app:EstimationPSD}. We note that all the PSDs in this article are given in one-side power spectral densities, so the frequency is only positive.
 
 To compute the Fourier transform of \ac{TDI} variables, we should consider the atomic block in \ac{TDI} formulation: the nested delay operator.
We assume the light travel times are constants, i.e., $L_{ij}(t) = L_{ij}$.
For a nested delay operator applied to a time series, $v(t) = \tdidelay{i_1 i_2 ... i_{n+1}} u(t)$, the Fourier transform is
 \begin{eqnarray}
 \tilde{v}(\omega) =  \exp \left(-j\omega\sum\limits_{k = 1}^{n} L_{i_k i_{k+1}}\right) \tilde{u}(\omega),\label{eq:Fourier_transform_nested_delay_fixed_arms}
\end{eqnarray}
with $n$ the number of delays.

The \ac{PSD}s of the usual \ac{TDI} generators ($X, Y$, and $Z$) are usually compositions of a limited set of patterns. 
For each term, we use \eqref{eq:Fourier_transform_nested_delay_fixed_arms} to form the Fourier transform and then compute the \ac{PSD}. 

We will use the shorthand notation 
\begin{eqnarray}
  \bar{L}_{ij} = \frac{L_{ij} + L_{ji}}{2} 
  \quad \textrm{and} \quad 
  \bar{L}_{ijik} = \frac{L_{ij} + L_{ji} + L_{ik} + L_{ki}}{4} \nonumber \\
\end{eqnarray}

\bgroup
\def\arraystretch{1.5}
\begin{table*}[ht]
   \centering
   \begin{tabular}{|c|c|}
    \hline
    Nested delay operator & PSD  \\
    \hline
    \hline
    $\pm \left(1 - \tdidelay{iji}\right)u(t)$ & $ 4\sin^2\left( \omega \bar{L}_{ij} \right) S_u $ \label{eq:PSDCal_terms_1a}\\
    $\pm \left(1 + \tdidelay{iji}\right)u(t)$ & $ 4\cos^2\left( \omega \bar{L}_{ij}  \right) S_u $ \label{eq:PSDCal_terms_1b}\\
    \hline
    $\pm \left(1 - \tdidelay{iji}\right) \tdidelay{i_1 i_2 ... i_n}u(t)$  & $ 4 \sin^2\left( \omega \bar{L}_{ij}  \right) S_u $ \label{eq:PSDCal_terms_2a}\\
    $\pm \left(1 + \tdidelay{iji}\right) \tdidelay{i_1 i_2 ... i_n}u(t)$  & $ 4 \cos^2\left( \omega \bar{L}_{ij}  \right) S_u $
    \label{eq:PSDCal_terms_2b}\\
    $\pm \tdidelay{i_1 i_2 ... i_n} \left(1 - \tdidelay{iji}\right)u(t)$  & $ 4 \sin^2\left( \omega \bar{L}_{ij}  \right) S_u $\\
    $\pm \tdidelay{i_1 i_2 ... i_n} \left(1 + \tdidelay{iji}\right)u(t)$  & $ 4 \cos^2\left( \omega \bar{L}_{ij}  \right) S_u $\\
    \hline
    $\pm\left(1 + \tdidelay{iji}\right)\left(1 - \tdidelay{klk}\right)u(t)$ & $ 16 \cos^2\left( \omega \bar{L}_{il} \right)\sin^2\left( \omega \bar{L}_{kl} \right) S_u $\label{eq:PSDCal_terms_3a} \\
    $\pm\left(1 - \tdidelay{iji}\right)\left(1 + \tdidelay{klk}\right)u(t)$ & $ 16 \sin^2\left( \omega \bar{L}_{ij} \right)\cos^2\left( \omega \bar{L}_{kl} \right) S_u $\label{eq:PSDCal_terms_3b} \\
    $\pm\left(1 + \tdidelay{iji}\right)\left(1 + \tdidelay{klk}\right)u(t)$ & $ 16 \cos^2\left( \omega \bar{L}_{ij} \right)\cos^2\left( \omega \bar{L}_{kl} \right) S_u $ \label{eq:PSDCal_terms_3c} \\
    $\pm\left(1 - \tdidelay{iji}\right)\left(1 - \tdidelay{klk'}\right)u(t)$ & $ 16 \sin^2\left( \omega \bar{L}_{ij} \right)\sin^2\left( \omega \bar{L}_{kl} \right) S_u $ \label{eq:PSDCal_terms_3d} \\
    \hline
    $\pm\left( 1 - \tdidelay{iji} - \tdidelay{ijiki} + \tdidelay{ikijiji}\right) u(t)$  & $ 16 \sin^2\left(\omega \bar{L}_{ij}\right) \sin^2\left(2 \omega \bar{L}_{ijk}\right)S_u $ \label{eq:PSDCal_terms_4} \\
    \hline
    $(a \pm b\tdidelay{iji}) x (t)$ & $ \left[a^2 + b^2 \pm 2ab \cos \left( \omega \bar{L}_{ij} \right) \right] S_u $ \label{eq:PSDCal_terms_5} \\
    \hline
   \end{tabular}
   \caption{Table of \ac{PSD} for the usual patterns present in \ac{TDI} time domain formulations.}
   \label{tab:nested_patterns}
   \end{table*}
   
Here, the PSD computation is done for the simple nested delay operator $\pm \left(1 - \tdidelay{i ji}\right)u(t)$.
The list of all useful patterns is provided in Table~\ref{tab:nested_patterns}.
 \begin{eqnarray} 
 {\rm PSD} & & \left[\pm\left(1 - \tdidelay{iji}\right)u(t)\right](\omega) \nonumber\\	& & = 
	\left\langle \widetilde{\left[\left(1 - \tdidelay{iji}\right)u(t)\right]}(\omega)\times
	\widetilde{\left[\left(1 - \tdidelay{iji}\right)u(t)\right]}^*(\omega)
	\right\rangle\nonumber\\
	& & = 
	\left\langle \left(1 - e^{-j\omega (L_{ij} + L_{ji})}\right)\left(1 - e^{j\omega (L_{ij} + L_{ji})}\right)\widetilde{u}(\omega)\widetilde{u}^*(\omega)
	\right\rangle\nonumber\\
	& & =
	4\sin^2\left( \omega \bar{L}_{ij} \right) S_u .
 \end{eqnarray}

 The \ac{CSD} computation have some common patterns. Note that we need to respect the order of the terms in the calculation. 
 \begin{enumerate}
 \itemsep0em
 \item $X = \pm (1 \pm \tdidelay{iji}) x (t)$ and $Y = \pm \left(1 \pm \tdidelay{klk}\right)u(t)$. We choose one case of specific set of signs in front of the nested delay operators, the others are easily worked out in the same way.
 \begin{eqnarray}
	CSD\left[X,Y\right] & = & CSD \left[(1 - \tdidelay{iji}) u(t), \left(1 + \tdidelay{klk}\right)u(t) \right] \nonumber \\
	& = & \left\langle \widetilde{\left[\left(1 - \tdidelay{iji}\right)u(t)\right]}(\omega)\times
	\widetilde{\left[\left(1 + \tdidelay{klk}\right)u(t)\right]}^*(\omega)\right\rangle\nonumber\\
	& = & \left\langle \left(1 - e^{-2 j \omega \bar{L}_{ij}} \right)\left(1 + e^{2 j \omega \bar{L}_{kl}}\right) \times \widetilde{u}(\omega) \widetilde{u}^*(\omega)
	\right\rangle\nonumber\\ 
	& = & e^{j\omega (- \bar{L}_{ij} + \bar{L}_{kl})} \left(e^{j\omega \bar{L}_{ij}} - e^{-j\omega \bar{L}_{ij}}\right) \nonumber \\
	& & \ \times \left(e^{-j\omega \bar{L}_{kl}} + e^{j\omega \bar{L}_{kl}}\right)\left\langle \widetilde{u}(\omega)\widetilde{u}^*(\omega)
	\right\rangle\nonumber\\
	& = & e^{j\omega (- \bar{L}_{ij} + \bar{L}_{kl})}  2j \sin (\omega\bar{L}_{ij})  2j \cos (\omega\bar{L}_{kl}) S_u \nonumber \\
	& = & - 4 \sin (\omega\bar{L}_{ij}) \cos (\omega\bar{L}_{kl}) e^{j\omega (- \bar{L}_{ij} + \bar{L}_{kl})}  S_u \label{eq:CSDCal_terms_1}
 \end{eqnarray}
 \item $X = \pm (a \pm b\tdidelay{iji}) x (t)$ and \\$Y = \pm \left(1 \pm \tdidelay{klk}\right)\tdidelay{i_1 i_2 ... i_n}u(t)$. We choose one case of specific set of signs in front of the nested delay operators, the others are easily worked out in the same way.
 \begin{eqnarray}
	CSD\left[X,Y\right] & = & CSD \left[(a + b\tdidelay{iji}) u(t) * \left(1 - \tdidelay{klk}\right)\tdidelay{i_1 i_2 ... i_n} u(t) \right] \nonumber \\
	& = & \left\langle \widetilde{\left[\left(a + b\tdidelay{iji}\right)u(t)\right]}(\omega)\right. \nonumber \\
	& & \ \left. \times
	\widetilde{\left[\left(1 - \tdidelay{klk}\right)\tdidelay{i_1 i_2 ... i_n}u(t)\right]}^*(\omega)\right\rangle\nonumber\\
	& = & \left\langle \left(a + b e^{-j\omega (L_{ij} + L_{ji})}\right)\left(1 - e^{j\omega (L_{kl} + L_{lk})}\right) \right. \nonumber \\
	& & \ \times \left. e^{j\omega (L_{i_1} + L_{i_2} + \ldots + L_{i_n})}\widetilde{u}(\omega)\widetilde{u}^*(\omega)
	\right\rangle\nonumber\\ 
	& = & e^{j\omega (L_{i_1} + L_{i_2} + \ldots + L_{i_n} - \bar{L}_{ij} + \bar{L}_{kl})} \left(e^{-j\omega \bar{L}_{kl}} - e^{j\omega \bar{L}_{kl}}\right) \nonumber \\
	& & \ \times \left(a e^{j\omega \bar{L}_{ij}} + be^{-j\omega \bar{L}_{ij}}\right) \left\langle \widetilde{u}(\omega)\widetilde{u}^*(\omega)
	\right\rangle\nonumber\\
	& = & -2j \sin (\omega\bar{L}_{kl}) e^{j\omega (L_{i_1} + L_{i_2} + \ldots + L_{i_n} - \bar{L}_{ij} + \bar{L}_{kl})} \nonumber \\
	& & \ \times \left(a e^{j\omega \bar{L}_{ij}} + be^{-j\omega \bar{L}_{ij}}\right) S_u  . \label{eq:CSDCal_terms_2}
 \end{eqnarray}
 
\end{enumerate}

 \subsection{Approximation justification} \label{sec:method:Approximation}
 In the previous subsections, some assumptions and approximations are made to reduce the complexity of the calculation. 
They are collected and justified here.  

 \begin{enumerate}
 \itemsep0em
 \item We assume that clock noise has been suppressed totally by the clock noise reduction algorithm~\cite{Hartwig:2020tdu}. 
 Therefore we do not need to consider the sideband beams in our beam model, since they are only used for clock noise reduction. Since the residual clock noise is expected below secondary noises, this assumption is acceptable in our study case. 
 \item All measurements are perfectly synchronized in the barycentric coordinate time. Hence, there are no errors in time stamping the on-board measurements. This assumption simplifies the complexity of the computation.
 \item All interferometric measurements are expressed as fractional frequency fluctuations around the nominal laser frequency. We assume this nominal laser frequency is constant and equal for all laser source, and it is equal to the nominal laser frequency, $c/1064 \ \textrm{nm} = 282 \ \textrm{THz}$. 
 \item The \ac{DFACS} is ignored in this study, which means the \ac{S/C} and test-masses are treated as independent bodies. We also neglect the tilt-to-length coupling noise in the beam model.
 \item  We are assuming that \ac{S/C} hardware from the noise performance perspective are statistically identical. Hence 6 test-mass acceleration noises have the same \ac{PSD}, or a correlation noise appearing between two adjacent test-masses will occur similarly on all \ac{S/C}.
 \item All armlengths of the LISA constellation are constant, and so delay operators are commutative. We use this approximation frequently with unsuppressed noises because the armlength variation is a second-order effect for these noises. 
 Therefore, this approximation is justified in the study of unsuppressed noises.
 \begin{equation}
 L_{ij}(t) = L_{ij} \qquad \forall i,j \in \{1,2,3\} \label{eq:approx:ConstantArms}
 \end{equation}
 \item Mostly in the case of unsuppressed noises, we neglect ranging and interpolation errors so the propagation delay operators and the \ac{TDI} delay operators can be treated similarly, $\propdelay{} \approx \tdidelay{}$. The effect of ranging and interpolation errors will contribute more significantly in the case of suppressed noises but this is out of the scope of this article.
 \item To simplify the final transfer functions, we use the approximation of equal armlengths, which could be considered as the average armlength for long duration of the mission operation. Due to the almost equilateral configuration of the LISA constellation, we expect the average of each armlength should be not too different.
 \end{enumerate}
 
In the simulation validation studies (see Sec.~\ref{sec:ValidationSimulation}), the 5 first approximations (no clock jitter noise, synchronized measurements, constant nominal laser frequency, no DFACS and noises of the same kind statistically similar) are made.
The validity of these approximations will not be tested here, whereas it will be for approximations 6 to 8.

 \subsection{Procedure for spectral density computation}\label{sec:psd_procedure}
 
 We will now detail the calculation of the transfer functions for unsuppressed noises, using as example test-mass acceleration noise. The propagation of other unsuppressed noises are worked out in a similar way.

 The calculations are performed in several steps:
 \begin{enumerate}
 \item If we consider laser frequency planning, laser noises from the locking scheme should be substituted into the beam model\footnote{An alternative approach is shown in Sec.~12.2 of~\cite{PhDHartwig}. In principle, TDI makes sure all the $p_{ij}$ terms are strongly suppressed, so any secondary noise terms in $p_{ij}$ due to laser locking are suppressed alongside the laser noise. Therefore, we expect the secondary noise levels to remain identical regardless of the locking scheme, as verified by the explicit computation.}.
 \item Since most of the time, we assume that noises of different types are uncorrelated, we can ignore all noises in the beams except for the one of interest. The \ac{LISA} total noise transfer function is then simply the sum of all individual noise transfer functions. If a noise correlation scenario is considered, we need to apply the correlation relations and keep only one of the correlated noises in the beam model.
 \item Next step is the computation of \ac{TDI} variables, presented in Sec.~\ref{sec:lisa-model:tdi}. First are the intermediary variables, then the TDI combinations. We write the result in terms of the product of nested delay operator applied to each noise, to ease the identification of patterns in the next step.
 \item Hence, we can use the patterns PSD/CSD presented in Sec.~\ref{sec:method:PsdCsdComputation} for quick computation of the spectral density of individual noise terms. The noise terms are considered uncorrelated. The correlations are treated by introducing the same noise term in multiple measurements. 
 \item We use the approximation of constant armlengths \eqref{eq:approx:ConstantArms} to simplify the computation (allowing to commute delay operators). Most of the time, the PSD $XX$ and the CSD $XY$ are enough because we can use index permutation to deduce the other spectral densities. This apply if the beams are symmetric, so it does not for the cases with frequency planning. 
 \item Finally, we sum up all components and simplify the result using some approximations presented in the end of Sec.~\ref{sec:method:Approximation}.
 \end{enumerate}
 
 \subsection{A few examples}
 \label{sec:examples}
 
 \subsubsection{Uncorrelated test-mass acceleration noise without laser locking}
 \label{sec:example_tm_uncorr_nolocking}

In this section, we only consider test-mass acceleration noise.
For simplicity, we omit the time dependency in the noise notation $\delta$, 
but still remember that it is a time varying signal.
We only consider the projection of test-mass displacement noise on the sensitive axis, $\delta_{ij}$, since it is what enters the measurements.

 Without frequency planning and correlation, the formulation of the measurements in \ac{S/C} 1 are:
 \begin{equation}
 \left\{
 \begin{array}{lll}
    \isig{12}  & = & 0\\
    \rfig{12}  & = & 0 \\
    \tmig{12}  & = &  2 \ \filter \ \theta_{12}^{\rfi} \ \delta_{12}
 \end{array}
 \right.
 \quad
 \left\{
 \begin{array}{lll}
    \isig{13}  & = &  0 \\
    \rfig{13} & = & 0 \\
    \tmig{13} & = & 2 \ \filter \ \theta_{13}^{\rfi} \ \delta_{13} 
 \end{array}
 \right.
 \end{equation}
 We then compute the TDI intermediary variables. 
 We neglect the ranging and interpolation errors such that the 
 two types of delay operators are equivalent, 
 $\propdelay{} \approx \tdidelay{}$. 
 Moreover, the nominal laser wavelength for every laser source is constant and equal, i.e., $\lambda_{ij} = \lambda$.
Applying these approximation to Eqs.~\eqref{eq:tdi_xi12},~\eqref{eq:tdi_xi13},~\eqref{eq:tdi_eta12} and ~\eqref{eq:tdi_eta13}, we get
 \begin{eqnarray}
  \xi_{12} 
  & = & - \theta_{12}^{\isi} \ \filter \left( \tdidelay{12} \delta_{21} + \delta_{12} \right),\\ 
  \xi_{13} 
  & = & - \theta_{13}^{\isi} \ \filter \left(\tdidelay{13} \delta_{31} + \delta_{13} \right),
 \end{eqnarray}
 and then
 \begin{eqnarray}
  \eta_{12} 
  		& = & - \filter \left(\tdidelay{12} \delta_{21} + \delta_{12} \right), \\
  \eta_{13} 
  		& = & - \filter \left(\tdidelay{13} \delta_{31} +  \delta_{13} \right).
\end{eqnarray}

The Michelson combination is computed as follows, using the constant armlength approximation \eqref{eq:approx:ConstantArms} (we can commute the delay operators with themselves and with antialiasing filter operator\footnote{This is not true in the case of suppressed noises like laser frequency noise. In such cases, we need to take into account the noncommutation of delay operators with themselves and with filter operators~\cite{Bayle:2018hnm}.}).
\begin{eqnarray}
 X_{2} & = & 
        \left(1 - \tdidelay{12131}\right)
        \left[\left(
        \eta_{13} + \tdidelay{13} \eta_{31} \right)
        \right. \nonumber \\
    & & \left. +\tdidelay{131}\left(\eta_{12} + \tdidelay{12} 
        \eta_{21} \right)\right]          
        - \left(1 - \tdidelay{13121}\right) \nonumber \\
    & & \times
        \left[\left(
        \eta_{12} + \tdidelay{12} \eta_{21} \right)+
        \tdidelay{121}\left(\eta_{13} + \tdidelay{13} 
        \eta_{31} \right)\right] \nonumber \\
    & \approx & \left(1 - \tdidelay{12131}\right) \left[ \left(1 - \tdidelay{121}\right)\left(
        \eta_{13} + \tdidelay{13} \eta_{31} \right) \right. \nonumber \\
        & & \qquad \qquad \left. - \left(1 - \tdidelay{131}\right)\left(
        \eta_{12} + \tdidelay{12} \eta_{21} \right)\right] \nonumber \\
        & = & \filter \bigg\{ - \left(1 - \tdidelay{12131}\right) \left(1 - \tdidelay{121}\right)\left(1 + \tdidelay{131}\right) \delta_{13} \nonumber \\
        & & \qquad - 2 \left(1 - \tdidelay{12131}\right) \left(1 - \tdidelay{121}\right) \tdidelay{13} \delta_{31} \nonumber \\
        & & \qquad + \left(1 - \tdidelay{12131}\right) \left(1 - \tdidelay{131}\right)\left(1 + \tdidelay{121}\right) \delta_{12} \nonumber \\
        & & \qquad + 2  \left(1 - \tdidelay{12131}\right) \left(1 - \tdidelay{131}\right) \tdidelay{12} \delta_{21}\bigg\}
 \end{eqnarray}
 The $Y$-channel is just the index permutation of $X$-channel.
 \begin{eqnarray}
 Y_{2} & = & 
        \filter \bigg\{ - \left(1 -       \tdidelay{23212}\right) \left(1 - \tdidelay{232}\right)\left(1 + \tdidelay{212}\right) \delta_{21} \nonumber \\
        & & \qquad - 2 \left(1 - \tdidelay{23212}\right) \left(1 - \tdidelay{232}\right) \tdidelay{21} \delta_{12} \nonumber \\
        & & \qquad + \left(1 - \tdidelay{23212}\right) \left(1 - \tdidelay{212}\right)\left(1 + \tdidelay{232}\right) \delta_{23} \nonumber \\
        & & \qquad + 2  \left(1 - \tdidelay{23212}\right) \left(1 - \tdidelay{212}\right) \tdidelay{23} \delta_{32}\bigg\}
 \end{eqnarray}
 
 The \ac{PSD} of these Michelson variables can be worked out by collecting the Fourier transforms of the auto-correlation functions of each noise in each \ac{MOSA}. 
Assuming uncorrelated noises, the cross-terms between two different noises, such as $\langle\widetilde{\delta_{12}}^*(f) \widetilde{\delta_{13}}(f)\rangle$, are vanishing. 
We can also use results from Sec.~\ref{sec:method:PsdCsdComputation} for fast deduction. For example, the contribution to the \ac{PSD} of $X$-channel $S_{XX}(f)$ of acceleration noise in \ac{MOSA} 13 reads:
 \begin{eqnarray}
 & & \textrm{PSD}\left[ - \filter \left(1 - \tdidelay{12131}\right) \left(1 - \tdidelay{121}\right)\left(1 + \tdidelay{131}\right) \delta_{13}\right] (\omega) \nonumber \\
 & & = 64 S_\mathcal{F}(\omega) S_{\delta_{13}}(\omega) \sin^2\left[\omega(\bar{L}_{12} + \bar{L}_{31})\right] \nonumber \\
 & & \quad \times  \sin^2(\omega \bar{L}_{12}) \cos^2(\omega \bar{L}_{31}) ,
 \end{eqnarray}
where $S_\mathcal{F}(\omega) = \langle|\tilde{\mathcal{F}}(f)|^2\rangle$ 
and
$S_{\delta_{13}}(\omega) = \langle|\widetilde{\delta_{13}}(f)|^2\rangle$.
Then, one can check that the \ac{PSD} of the $X$-channel for the uncorrelated test-mass acceleration noise is
 \begin{eqnarray}
 \textrm{S}_\textrm{XX}^\textrm{uncorr acc tm} (\omega) & = &   64 S_\mathcal{F} (\omega) \sin^2\left[\omega(\bar{L}_{12} + \bar{L}_{31}) \right] \nonumber \\
 & & \times \bigg\{ \sin^2 (\omega \bar{L}_{12}) \left[\cos^2 (\omega \bar{L}_{31})  S_{\delta_{13}}(\omega) \right. \nonumber \\
 & & \left. + S_{\delta_{31}}(\omega) \right] + \sin^2 (\omega \bar{L}_{31}) \nonumber \\
 & & \times \left[ \cos^2(\omega \bar{L}_{12}) S_{\delta_{12}}(\omega) + S_{\delta_{21}} (\omega) \right]\bigg\} \nonumber \\
 \end{eqnarray}
 
The \ac{PSD} of $Y$-channel, $\textrm{S}_\textrm{YY}^\textrm{uncorr acc tm}$, has the same form with permuted indices $\{1 \rightarrow 2, 2 \rightarrow 3, 3 \rightarrow 1\}$. 
We can use the equal armlength approximations $L_{ij} = L$ and that all test-mass acceleration noises share the same \ac{PSD}, $S_{\delta_{ij}} = S_\delta$, to get:
 \begin{eqnarray}
\textrm{S}_\textrm{XX}^\textrm{uncorr acc tm} (\omega) & = &  \textrm{S}_\textrm{YY}^\textrm{uncorr acc tm} (\omega) \nonumber \\
 & = &  64 \sin^2 \left(2 \omega L\right) \sin^2 \left(\omega L\right) \left[3 + \cos(2 \omega L)\right] \nonumber \\
 & & \qquad \qquad \qquad \qquad \times S_\mathcal{F} (\omega) S_\delta (\omega)
 \label{eq:tm_accel_XX_uncorr_no_locking}
 \end{eqnarray}

 To compute the \ac{CSD} between $X$ and $Y$, we use the same procedure and collect the nonzero terms that have the same noise index. Note that $CSD[Y,X] = CSD[X,Y]^*$, so we only need to compute the \ac{CSD} of $XY$. We can also use the \ac{CSD} result from Sec. \ref{sec:method:PsdCsdComputation}. For example, the contribution of acceleration noise in \ac{MOSA} 12 to the \ac{CSD} $S_{XY}$ reads:
 
 \begin{eqnarray}
 & & \textrm{CSD}\bigg[ \filter \left(1 - \tdidelay{12131}\right) \left(1 - \tdidelay{131}\right)\left(1 + \tdidelay{121}\right) \delta_{12}  \nonumber \\
 & & \qquad \quad  * (-2) \left(1 - \tdidelay{23212}\right) \left(1 - \tdidelay{232}\right) \tdidelay{21} \delta_{12}\bigg]  (\omega) \nonumber \\
 & = & - 64 S_\mathcal{F} (\omega) S_{\delta_{12}} (\omega) \sin\left[\omega(\bar{L}_{12} + \bar{L}_{31}) \right] \nonumber \\
 & & \times \sin\left[\omega(\bar{L}_{12} + \bar{L}_{23}) \right] \sin(\omega \bar{L}_{13}) \sin(\omega \bar{L}_{23}) \cos(\omega \bar{L}_{12}) \nonumber \\
 & & \times \exp\left[-j\omega \left(2 \bar{L}_{13} - 2 \bar{L}_{23} + \bar{L}_{12} - L_{21} \right) \right]
 \end{eqnarray}
 
 One can find the \ac{CSD} of $XY$ is given by
 \begin{eqnarray}
 \textrm{S}_\textrm{XY}^\textrm{uncorr acc tm} (\omega) 
 & = &  - 64 S_\mathcal{F} (\omega) \sin\left[\omega(\bar{L}_{12} + \bar{L}_{31}) \right] \nonumber \\
 & & \times \sin\left[\omega(\bar{L}_{12} + \bar{L}_{23}) \right] \sin(\omega \bar{L}_{13}) \nonumber \\
 & & \times \sin(\omega \bar{L}_{23}) \cos(\omega \bar{L}_{12}) 
 e^{-j \omega \frac{L_{12} - L_{21}}{2}} \nonumber \\
 & & \times e^{-2j\omega \left(\bar{L}_{13} - \bar{L}_{23} \right)}  \left[S_{\delta_{12}} (\omega) + S_{\delta_{21}} (\omega) \right] \nonumber \\
 \end{eqnarray}
 
 Assuming equal armlengths and the same test-mass acceleration noise level in all \ac{MOSAs}, we obtain
 \begin{eqnarray}
 \textrm{S}_\textrm{XY}^\textrm{uncorr acc tm} (\omega)
 & = &  - 64 S_\mathcal{F} (\omega) \sin^3\left(2 \omega L \right) \sin\left(\omega L \right) S_{\delta} (\omega) \nonumber \\
 \label{eq:tm_accel_XY_uncorr_no_locking}
 \end{eqnarray}

 \subsubsection{Uncorrelated test-mass acceleration noise with laser locking}
 \label{sec:example_tm_uncorr_locking}
 
 To account for frequency planning, we need to derive the locked laser frequency fluctuations as functions of the primary laser, $p_{32}$, before substituting them in the beam model and interferometric measurements.
We use the group of equations \eqref{eq:freq_plan_N2c} and we only keep track of the test-mass acceleration and primary laser noises,
 \begin{subequations}
 \begin{align}
   p_{23}    & = \propdelay{12} \ p_{32}\\
   p_{31}    & = p_{32}\\
   p_{13}    & = \propdelay{21} \ p_{32} \\
   p_{12}    & = \propdelay{21} \ p_{32} \\
   p_{21}    & = \propdelay{321} \ p_{32}.
 \end{align}
 \end{subequations}
 
 Due to laser locking, the beams and interferometric measurements are no longer symmetric for the different \ac{S/C}.
We therefore give the interferometric signals for the whole LISA constellation 
 \begin{enumerate}[\textbullet]
 \item On \ac{S/C} 1:
 \begin{eqnarray}
 & &\left\{
 \begin{array}{lll}
    \isig{12}  & = & \theta_{12}^{\isi} \filter \left(\propdelay{121} - 1\right) \propdelay{13}  p_{32} \\
    \rfig{12}  & = & 0 \\
    \tmig{12}  & = & 2 \ \filter \  \theta_{12}^{\rfi} \ \delta_{12}
 \end{array}
 \right.\\
 & &\left\{
 \begin{array}{lll}
    \isig{13}  & = & 0 \\
    \rfig{13}  & = & 0 \\
    \tmig{13}  & = & 2 \ \filter \ \theta_{13}^{\rfi} \ \delta_{13}
 \end{array}
 \right.
 \end{eqnarray}
 
 \item On \ac{S/C} 2:
 \begin{eqnarray}
 & &\left\{
 \begin{array}{lll}
    \isig{23}  & = & 0 \\
    \rfig{23}  & = & \theta_{23}^{\rfi} \filter \left(\propdelay{213} - \propdelay{23}\right) p_{32}\\
    \tmig{23}  & = & \theta_{23}^{\rfi} \filter \left[\left(\propdelay{213} - \propdelay{23}\right) p_{32} +  2 \delta_{23}\right]
 \end{array}
 \right.\\
 & &\left\{
 \begin{array}{lll}
    \isig{21}  & = & 0 \\
    \rfig{21}  & = & \theta_{21}^{\rfi} \left(\propdelay{23} - \propdelay{213}\right) p_{32} \\
    \tmig{21}  & = & \theta_{21}^{\rfi} \filter \left[\left(\propdelay{23} - \propdelay{213}\right) p_{32} +  2 \delta_{21}\right]
 \end{array}
 \right.
 \end{eqnarray}
 
 \item On \ac{S/C} 3:
 \begin{eqnarray}
 & &\left\{
 \begin{array}{lll}
    \isig{31}  & = & \theta_{31}^{\isi} \left(\propdelay{313} - 1\right)p_{32} \\
    \rfig{31}  & = & 0 \\
    \tmig{31}  & = & 2 \filter \theta_{31}^{\rfi} \delta_{31}
 \end{array}
 \right.\\
 & &\left\{
 \begin{array}{lll}
    \isig{32}  & = & \theta_{32}^{\isi} \left(\propdelay{323} - 1\right)p_{32} \\
    \rfig{32}  & = & 0 \\
    \tmig{32}  & = & 2 \filter \theta_{32}^{\rfi} \delta_{32}
 \end{array}
 \right.
 \end{eqnarray}
 \end{enumerate}
 
 The next step is to compute the \ac{TDI} intermediary variables $\xi, \eta$. Assuming $\propdelay{} = \tdidelay{}$, one can verify that
 \begin{eqnarray}
    \eta_{12} & = & \filter (\tdidelay{123} - \tdidelay{13})p_{32} -  \filter \left(\tdidelay{12} \delta_{21} + \delta_{12} \right) \\
  	\eta_{13} & = & -  \filter \left(\tdidelay{13} \delta_{31} +  \delta_{13} \right) \\ 
  	\eta_{23} & = &  -  \filter \left(\tdidelay{23} \delta_{32} + \delta_{23} \right) \\
  	\eta_{21} & = & \filter (\tdidelay{213} - \tdidelay{23})p_{32} - \filter \left(\tdidelay{21} \delta_{12} +  \delta_{21} \right) \\
  	\eta_{31} & = & \filter (\tdidelay{313} - 1)p_{32} -  \filter \left(\tdidelay{31} \delta_{13} + \delta_{31} \right) \\
  	\eta_{32} & = & \filter (\tdidelay{323} - 1)p_{32} -  \filter \left(\tdidelay{32} \delta_{23} +  \delta_{32} \right)
 \end{eqnarray}
 We note that, except for the terms with laser frequency noise $p_{32}$, all terms in $\eta$ are identical to the case  without laser locking. That is expected because the locking constraints \eqref{eq:freq_plan_N2c} do not contain test-mass acceleration noise in any term.
The $X$-channel for laser noise only is
 \begin{eqnarray}
    X_{2}^\text{p-only} & = & \filter \left[ (1 - \tdidelay{13121})(1 - \tdidelay{12131}) \right.\nonumber \\
    & & \quad \left. -  (1 - \tdidelay{12131})(1 - \tdidelay{13121}) \right] p_{32},
 \end{eqnarray}
 which is canceled out when we commute the \ac{TDI} delay, i.e., using constant delays assumption. 
In the end, the \ac{TDI} combinations $X$, $Y$ and $Z$ in the case of laser locking for the test-mass acceleration noise are exactly the same as in the case without laser locking, \eqref{eq:tm_accel_XX_uncorr_no_locking} and  \eqref{eq:tm_accel_XY_uncorr_no_locking}.

\subsubsection{Uncorrelated readout and optical path noises with laser locking}
\label{sec:example_ro_uncorr_locking}
 
The locking constraints \eqref{eq:freq_plan_N2c} contain readout noises, $N^{ro}_{x,ij}$, and optical path noises, $N^{op}_{loc/x,ij}$.
Therefore, the situation is different from acceleration noise. 
Expanding $\eta_{12}$ without laser locking, we get:

 \begin{eqnarray}
	\eta_{12} & = & \theta_{21}^{\isi} \filter N^{ro}_{s,12}
  		- \theta_{21}^{\rfi} \filter \tdidelay{12} \dfrac{N^{ro}_{\epsilon,21} - N^{ro}_{\rfi,21}}{2}  \nonumber\\
  		& & - \theta_{12}^{\rfi} \filter \dfrac{N^{ro}_{\epsilon,12} - N^{ro}_{\rfi,12}}{2}
  		+ \theta_{21}^{\rfi} \tdidelay{12} \filter \dfrac{N^{ro}_{\rfi,21} + N^{ro}_{\rfi,23}}{2}, \nonumber \\
\end{eqnarray}

while we get with laser locking:
 \begin{eqnarray}
    \eta_{12} & = &  \theta_{12}^{\isi} \filter N^{ro}_{s,12} - \theta_{21}^{\rfi} \tdidelay{12} \filter \frac{N^{ro}_{\epsilon,21} -  N^{ro}_{\rfi, 21}}{2}  \nonumber \\
	& & - \theta_{12}^{\rfi} \filter \frac{N^{ro}_{\epsilon,12} - N^{ro}_{\rfi,12}}{2}  + \theta_{21}^{\rfi} \tdidelay{12} \filter \frac{N^{ro}_{\rfi,21} + N^{ro}_{\rfi,23}}{2} \nonumber \\
	& & - \theta_{13}^{\isi} \filter N^{ro}_{s,13} + \theta_{23}^{\isi} \filter \tdidelay{12} N^{ro}_{s,23} \nonumber \\ 
	& & - \theta_{31}^{\rfi} \filter \tdidelay{13} N^{ro}_{\rfi, 31} - \theta_{12}^{\rfi} \filter N^{ro}_{\rfi,12}
 \end{eqnarray}
We observe that laser locking introduces additional terms. 
These terms actually vanish at the next \ac{TDI} step, when forming the variable $\eta$.
Considering, for example, solely $N^{ro}_{s,13}$, we have
  \begin{eqnarray*}
    \eta_{12} & = & - \theta_{13}^{\isi} N^{ro}_{s,13} , \\
    \eta_{21} & = & \theta_{13}^{\isi} \tdidelay{21} N^{ro}_{s,13} ,\\
    \eta_{31} & = & \theta_{13}^{\isi} \tdidelay{31} N^{ro}_{s,13} 
  \end{eqnarray*}
 Substituting in $X_{2}$ given by Eq.~\eqref{eq:TDIX20}, we get
 \begin{eqnarray}
   X_{2} & = & \left[ 1 - \tdidelay{121} -  \tdidelay{12131} + \tdidelay{1312121} 
   + \left( \tdidelay{13121} - \tdidelay{12131} \right) \right. \nonumber \\
   & & \left.  + \left( \tdidelay{131212131} - \tdidelay{121313121} \right)  \right] 
   \theta_{13}^{\isi} N^{ro}_{s,13} .
 \end{eqnarray}
 Assuming that delay operators commute, the terms in parentheses disappear and we are back to the results without laser locking. 
 
 One can checked that we obtain the same results as for the case without laser locking, for all terms of readout noises and  optical path noises. 
 Finally, we find that the results are the same with and without laser locking for all unsuppressed noises.

\subsubsection{Correlated acceleration noise}
\label{sec:example_tm_corr}
 
 Finally, we consider the correlation scenario \eqref{eq:tm_corr_scen} 
 for test-mass acceleration noise. The correlation relation is
 \begin{equation}
 \delta_{ij} = \gamma \ \delta_{ik} ,
 \end{equation}
 for $(i,j,k) = \text{circular permutation of} \ (1,2,3)$, with $\gamma$ the correlation factor and with $j \neq k$. We substitute this in the beam model and then form the interferometric measurements. Since the correlated noises are in the same \ac{S/C}, the interferometric measurements remain symmetric (as in the uncorrelated noise case). In \ac{S/C} 1, we keep only the test-mass acceleration noise from \ac{MOSA} 12, 
 
 \begin{equation}
 \left\{
 \begin{array}{lll}
    \isig{12}  & = & 0\\
    \rfig{12}  & = & 0 \\
    \tmig{12}  & = &  2 \ \filter \ \theta_{12}^{\rfi} \ \delta_{12}
 \end{array}
 \right.
 \quad
 \left\{
 \begin{array}{lll}
    \isig{13}  & = &  0 \\
    \rfig{13} & = & 0 \\
    \tmig{13} & = & 2 \ \filter \ \theta_{13}^{\rfi} \ \gamma \delta_{12} 
 \end{array}
 \right.
 \end{equation}

Then, the \ac{TDI} intermediary variables $\eta$ for \ac{S/C} 1 are
\begin{eqnarray}
    \eta_{12} & = &  -  \filter \left(\gamma \tdidelay{12} \delta_{23} + \delta_{12} \right), \\
  	\eta_{13} & = & -  \filter \left(\tdidelay{13} \delta_{31} + \gamma \delta_{12} \right)
 \end{eqnarray}

Applying the same procedure as for the uncorrelated case, we get the following expression for the \ac{PSD}:

\begin{eqnarray}
\textrm{S}_\textrm{XX}^\textrm{corr acc tm} (\omega)
 & = &  32 \left[3 \gamma^2 + 2 \gamma + 3 + \left( 1 + \gamma \right)^2 \cos(2 \omega L)\right]  \nonumber \\
 & & \times  \sin^2 \left(2 \omega L\right) \sin^2 \left(\omega L\right) S_\mathcal{F}(\omega) S_\delta(\omega),  \nonumber \\
 \label{eq:tm_accel_XX_corr_no_locking}
\end{eqnarray}
and, for the \ac{CSD},
\begin{eqnarray}
\textrm{S}_\textrm{XY}^\textrm{corr acc tm} (\omega)
 & = &  - 64 \left[ \left( 1 + \gamma \right)^2 \cos(2 \omega L) - \gamma \right]  \nonumber \\
 & & \times  \sin^2 \left(2 \omega L\right) \sin^2 \left(\omega L\right) S_\mathcal{F}(\omega) S_\delta(\omega)  \nonumber \\
 \label{eq:tm_accel_XY_corr_no_locking}
 \end{eqnarray}

This example is a good illustration of the importance of correlation. 
Indeed, at low frequency, $\cos(2 \omega L) \sim 1$, and the fully correlated case ($\gamma = 1$) is 1.5 times higher than the uncorrelated case. On the other hand, the fully anticorrelated case ($\gamma = -1$) case is 2 times lower than the uncorrelated case.

 \subsection{Validation with simulation}
 \label{sec:ValidationSimulation}
 
   \subsubsection{LISANode}

   \texttt{LISANode}~\cite{LISANodeSoftware} is the current official simulator of the LISA Consortium. 
   It is a time domain simulator based on a modular structure using graphs to connect blocks and finally core components. The core components are coded in \texttt{C++} and the rest (organization of components, graph building and validation, user interface) is in \texttt{Python}. Part of the logic and several elements are inherited from the \texttt{LISACode} simulator \cite{Petiteau:2006me, PhDPetiteau}.
   \texttt{LISANode} takes as inputs an orbit file, a frequency plan and potentially \ac{GW} files and glitch files. 
   It simulates the noises sources, the propagation of laser beams, the interferometric measurements, the phasemeters, the clocks, etc. It produces the interferometric measurements at 16 Hz. These measurements are then filtered and downsampled at 4 Hz to produce the telemetered data. The simulator is then connected to a processing module to apply TDI and produce any TDI variables. It has already been used in multiple studies~\cite{Bayle:2018hnm, Hartwig:2020tdu, Bayle:2021mue, Hartwig:2022yqw} and is described in \cite{PhDBayle, PhDHartwig}.

   \subsubsection{Numerical method for spectral estimation}
   
 The procedure to validate the transfer function of a particular type of noise (for example acceleration noise or readout noise) is the following:
 \begin{enumerate}[i)]
 \itemsep0em
 \item We configure the simulation for the noise to be studied, with all other noises configured to produce zeros as output;
 \item From the simulated time domain data, we compute the \ac{PSD} and the \ac{CSD};
 \item For the same set of frequencies, we compute data from our analytical formulation;
 \item We overplot the simulated and analytical \ac{PSD}s or \ac{CSD}s, adding for the analytical curve, the 99.73\% confidence interval (3~$\sigma$ for normal distribution) which is computed statistically for our Welch PSD/CSD estimates;
 \item The simulated points outside the confidence interval are detected.
 The level of agreement between analytical formulation and simulated data is estimated based on the plot and the number of points outside the confidence interval.
 \end{enumerate}

 \section{Results}
 \label{sec:results}
 \subsection{Propagation of unsuppressed noises}
 
  \subsubsection{Analytical formulations}
   To summarize all analytical results, we list the noises with the specific correlation
   and the \ac{TDI} transfer functions for $X$ in Table~\ref{tab:summary_unsup_table_XYZ}. 
   The results are the same for $Y$ and $Z$, even with laser locking. 
For all these results, the equal armlengths and equal noise level approximations are used. 
We do not distinguish between the case with or without laser locking, since the results are identical for the unsuppressed noises.
For the sake of brevity, we introduce two common factors in the summary table:
   \begin{eqnarray}
   \cxx &=& 16 \sin^2(\omega L) \sin^2(2 \omega L), \\
   \cxy &=& -16 \sin(\omega L) \sin^3(2 \omega L).
   \end{eqnarray}
   
  \bgroup
   \def\arraystretch{1.5}
   \begin{table*}[ht]
   \centering
   \begin{tabular}{|c|c|c|c|}
    \hline
    Noise type  & Correlation & PSD  & CSD \\
    \hline \hline
    \multirow{3}{3cm}{\centering test-mass acceleration} & None & $4 \cxx \left[3 + \cos(2 \omega L)\right]  $ & $4 \cxy $ \\
    & Correlated noises at the same \ac{S/C} & $8 \cxx$ & $ - 4 \cxx$ \\
    & Anticorrelated noises at the same \ac{S/C} & $8 \cxx \left[2 + \cos(2 \omega L)\right]$ & $4 \cxx \left[1 - 4 \cos(\omega L)\right]$ \\
    \hline
    \multirow{3}{3cm}{Readout (\ac{TMI}) and Optical Pathlength (\ac{TMI})} & None & $\cxx \left[3 + \cos(2 \omega L)\right]$ & $\cxy$ \\
    & Correlated adjacent TMI noise & $2 \cxx$ & $ - \cxx$ \\
    & Anticorrelated adjacent TMI noise & $ 2 \cxx \left[2 + \cos(2 \omega L)\right]$ & $\cxx \left[1 - 4 \cos(\omega L)\right]$ \\
    \hline
    Backlink (\ac{TMI}) & None & $\cxx \left[3 + \cos(2 \omega L)\right]$ & $\cxy$ \\
    \hline
    \multirow{4}{3cm}{\centering Readout (\ac{ISI} and \ac{RFI}) and Optical Pathlength (\ac{ISI} and \ac{RFI})} & None & $4 \cxx $ & $\cxy$ \\
    & Correlated adjacent IFO noise & $2 \cxx$ & $- \cxx$ \\
    & Anticorrelated adjacent IFO noise & $ 6 \cxx $ & $\cxx \left[1 - 4 \cos(\omega L)\right]$ \\
    & Correlated noises at the same telescope & $4 \cxx \left[3 + \cos(2 \omega L)\right]$ & $4\cxy$ \\
    \hline
    Backlink (\ac{RFI}) & None & $4 \cxx$ & $\cxy$ \\
    \hline
    
   \end{tabular}
   \caption{Summary table of analytical \ac{TDI} $X$,$Y$,$Z$ transfer functions for unsuppressed noises. All results have been simplified using approximations (refer to Sec.~\ref{sec:method:Approximation}).}
   \label{tab:summary_unsup_table_XYZ}
   \end{table*}
    
   Several types of noises share the same transfer function. 
   For some of them, it is simply because the noises enter identically in the measurement (e.g., readout \ac{ISI} and optical path \ac{ISI}). 
   
   As a result, we recover the transfer functions of all common noises existing in LISA sensitivity~\cite{SciRD} after using the approximations. New transfer functions of the optical path and readout noises in test-mass interferometers are found, which can change the shape of LISA sensitivity. In different noise correlation scenarios, the transfer functions could be either lower or higher than the ones in the uncorrelated case, which can help build the worst-case model for the LISA noise budget.
   
   There is another set of \ac{TDI} variables, called $A,E,T,$ constructed from $X,Y,Z$~\cite{Prince,Babak:2021mhe}: 
   \begin{eqnarray}
   A = \frac{Z-X}{\sqrt{2}}, \;
   E = \frac{X-2Y+Z}{\sqrt{6}}, 
   T = \frac{X+Y+Z}{\sqrt{3}}.
  \label{eq:AET}
  \end{eqnarray}
  $A,E,T$ are useful for data analysis since they have vanishing \ac{CSD}s under the approximations of equal armlengths and equal noise levels for the same type noises. 
  The \ac{PSD}s for $A,E,T$ are given in Table~\ref{tab:summary_unsup_table_AET}.
  They combine the \ac{PSD}s and \ac{CSD}s of $X,Y,Z$ as
  \begin{eqnarray}
  S_{AA} & = & \frac{S_{ZZ} + S_{XX} - 2 \textrm{Re} [ S_{ZX} ] }{2} \\
  S_{EE} & = & \frac{S_{XX} + 4 S_{YY} + S_{ZZ} - 2 \textrm{Re} [ 2 S_{XY} - S_{XZ} + 2 S_{YZ} ] }{6} \nonumber \\
  & & \\
  S_{TT} & = & \frac{S_{XX} + S_{YY} + S_{ZZ} + 2 \textrm{Re} [ S_{XY} + S_{XZ} + S_{YZ} ] }{3} \nonumber \\
\end{eqnarray} 
and are therefore slightly more complex.
  We remark that while the equal arm models derived here are accurate enough to describe the GW-sensitive channels $X,Y,Z,$ as well as for the quasiorthogonal channels $A$ and $E$, it was demonstrated that this assumption is insufficient for accurately describing the behavior of the null channel $T$, in particular at low frequencies~\cite{Muratore:2022nbh, AdamsCornish2010}.

  \bgroup
	\def\arraystretch{1.5}

   \begin{table*}[ht]
   \centering
   \begin{tabular}{|c|c|c|c|}
    \hline
    Noise type  & Correlation & PSD A \& E & PSD T \\
    \hline \hline
    \multirow{3}{3cm}{\centering test-mass acceleration} & None & $4 \cxx \left[3 + 2 \cos(\omega L) + \cos(2 \omega L)\right] $ & $ 32 \cxx  \sin^4(\frac{\omega L}{2}) $ \\
    & Correlated noises at the same \ac{S/C} & $12 \cxx $ & $0$ \\
    & Anticorrelated noises at the same \ac{S/C} & $4 \cxx \left[1 + 2 \cos(\omega L) \right]^2$ & $ 64 \cxx  \sin^4(\frac{\omega L}{2}) $ \\
    \hline
    \multirow{3}{3cm}{Readout (\ac{TMI}) and Optical Pathlength (\ac{TMI})} & None & $\cxx \left[3 + 2 \cos(\omega L) + \cos(2 \omega L)\right] $ & $ 8 \cxx  \sin^4(\frac{\omega L}{2}) $ \\
    & Correlated adjacent TMI noise & $3 \cxx$ & $0$\\
    & Anticorrelated adjacent TMI noise &  $\cxx \left[1 + 2 \cos(\omega L)\right]^2 $ & $16 \cxx \sin^4 \left( \frac{\omega L}{2} \right)$\\
    \hline
    Backlink (\ac{TMI}) & None & $\cxx \left[3 + 2 \cos(\omega L) + \cos(2 \omega L)\right] $ & $ 8 \cxx  \sin^4(\frac{\omega L}{2}) $ \\
    \hline
    \multirow{4}{3.5cm}{\centering Readout (\ac{ISI} and \ac{RFI}) and Optical Pathlength (\ac{ISI} and \ac{RFI})} & None & $ 2 \cxx \left[2 + \cos(\omega L) \right] $ & $4 \cxx \left[1 - \cos(\omega L) \right] $  \\
    & Correlated adjacent IFO noise & $3 \cxx$ & $0$\\
    & Anticorrelated adjacent IFO noise &  $\cxx \left[5 + 4 \cos(\omega L)\right] $ & $-8 \cxx \left[-1 + \cos(\omega L) \right]$\\
    & Correlated noises at the same telescope & $4 \cxx \left[3 + 2 \cos(\omega L) + \cos(2 \omega L)\right] $ & $ 32 \cxx  \sin^4(\frac{\omega L}{2}) $ \\
    \hline
    Backlink (\ac{RFI}) & None & $ 2 \cxx \left[2 + \cos(\omega L) \right] $ & $4 \cxx \left[1 - \cos(\omega L) \right] $ \\
    \hline
   \end{tabular}
   \caption{Summary table of analytical \ac{TDI} $A,E,T$ transfer functions for unsuppressed noises. All results have been simplified using approximations (refer to Sec.~\ref{sec:method:Approximation}).}
   \label{tab:summary_unsup_table_AET}
   \end{table*}

   \subsubsection{Analytic formulations versus simulations}
   
For the frequency range $10^{-4}$ to 1~Hz, the simulated and analytical \ac{PSD}/\ac{CSD} for TDI $X$ have been plotted (see Figs.~\ref{fig:validation_psd_accel_noise}, \ref{fig:validation_psd_accel_noise_corr}, \ref{fig:validation_psd_accel_noise_anti_corr} and \ref{fig:validation_csd_accel_noise}). 
Red lines show the analytical formulation expressions. The blue dashed lines represent the instrument response to the simulated single noises (i.e, the test-mass acceleration noise in the following example) for a duration about $3 \times 10^5$~s for uncorrelated and correlated cases and about $7 \times 10^4$~s for anticorrelated case. 
The green envelope highlights the 99.73\% confidence interval with respect to the analytical formulation. The width of the envelope depends on the confidence interval and on the duration of the simulation (see the difference between \ref{fig:validation_psd_accel_noise_anti_corr} and \ref{fig:validation_psd_accel_noise} and \ref{fig:validation_psd_accel_noise_corr}).
The probability that a single point is outside of the confidence interval is around 4.5$\times$10$^{-7}$ in case of a perfect agreement between analytical formulation and simulation (see appendix~\ref{app:EstimationPSD} and Eq.~\ref{eq:app_probConfInterv}).

Figures~\ref{fig:validation_psd_accel_noise}, \ref{fig:validation_psd_accel_noise_corr}, and \ref{fig:validation_psd_accel_noise_anti_corr} show a great agreement for the test-mass acceleration noise PSD in all uncorrelated, correlated and anticorrelated cases.

The confidence interval described in the appendix~\ref{app:EstimationPSD} is not applicable to the CSD. Nevertheless, the CSD computation shows good visual agreement with the simulated data from LISANode (see figure~\ref{fig:validation_csd_accel_noise}).

    \begin{figure}[h]
    \begin{center}
    \includegraphics[width=0.5\textwidth]{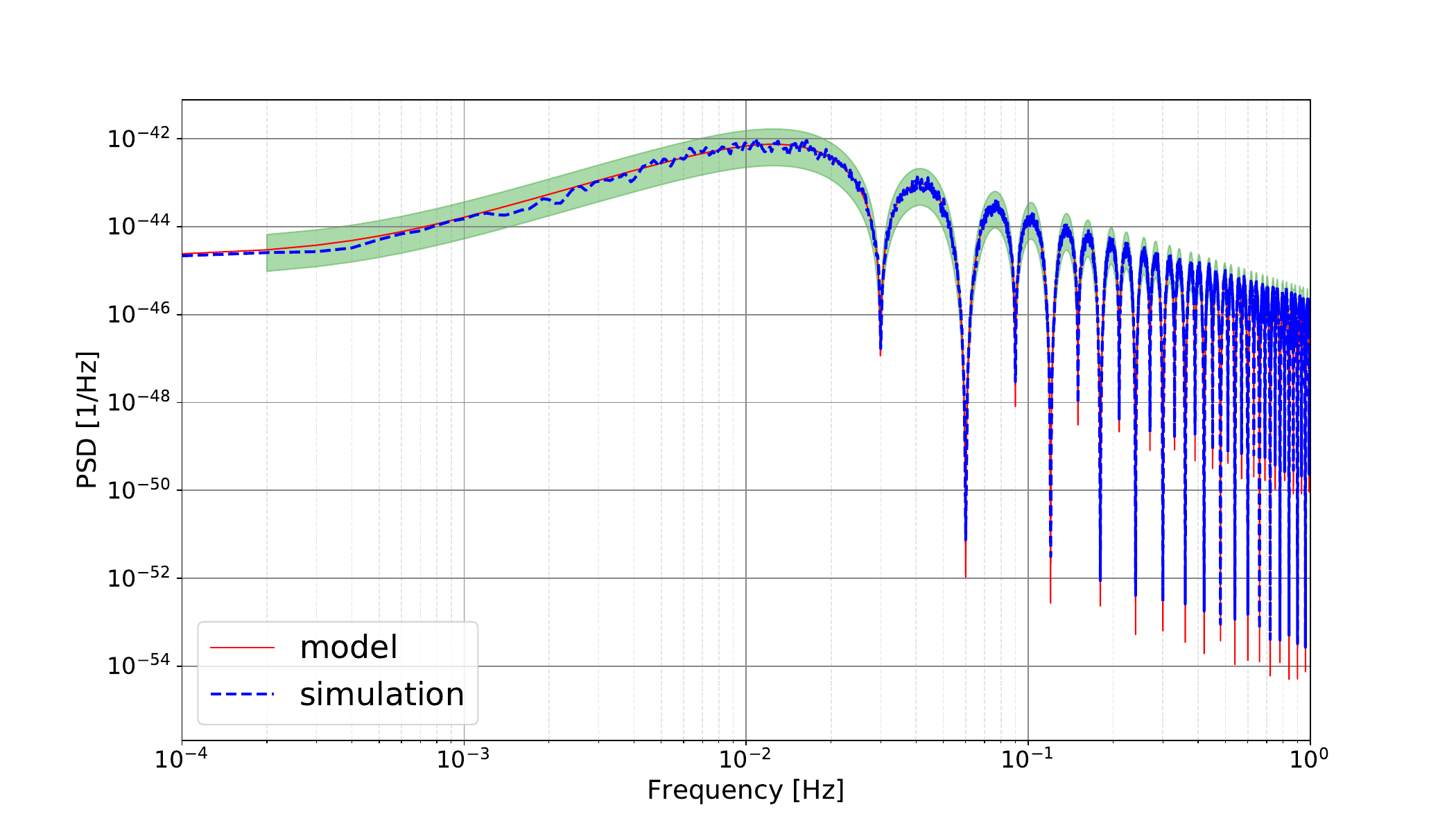}
    \caption{Uncorrelated test-mass acceleration noise cross-comparison. The simulated data (red line) at 99.73\% confidence interval (green area) are in great agreement with the analytical formulation (blue dashed line).}
    \label{fig:validation_psd_accel_noise}
    \end{center}
    \end{figure}

    \begin{figure}[h]
    \begin{center}
    \includegraphics[width=0.5\textwidth]{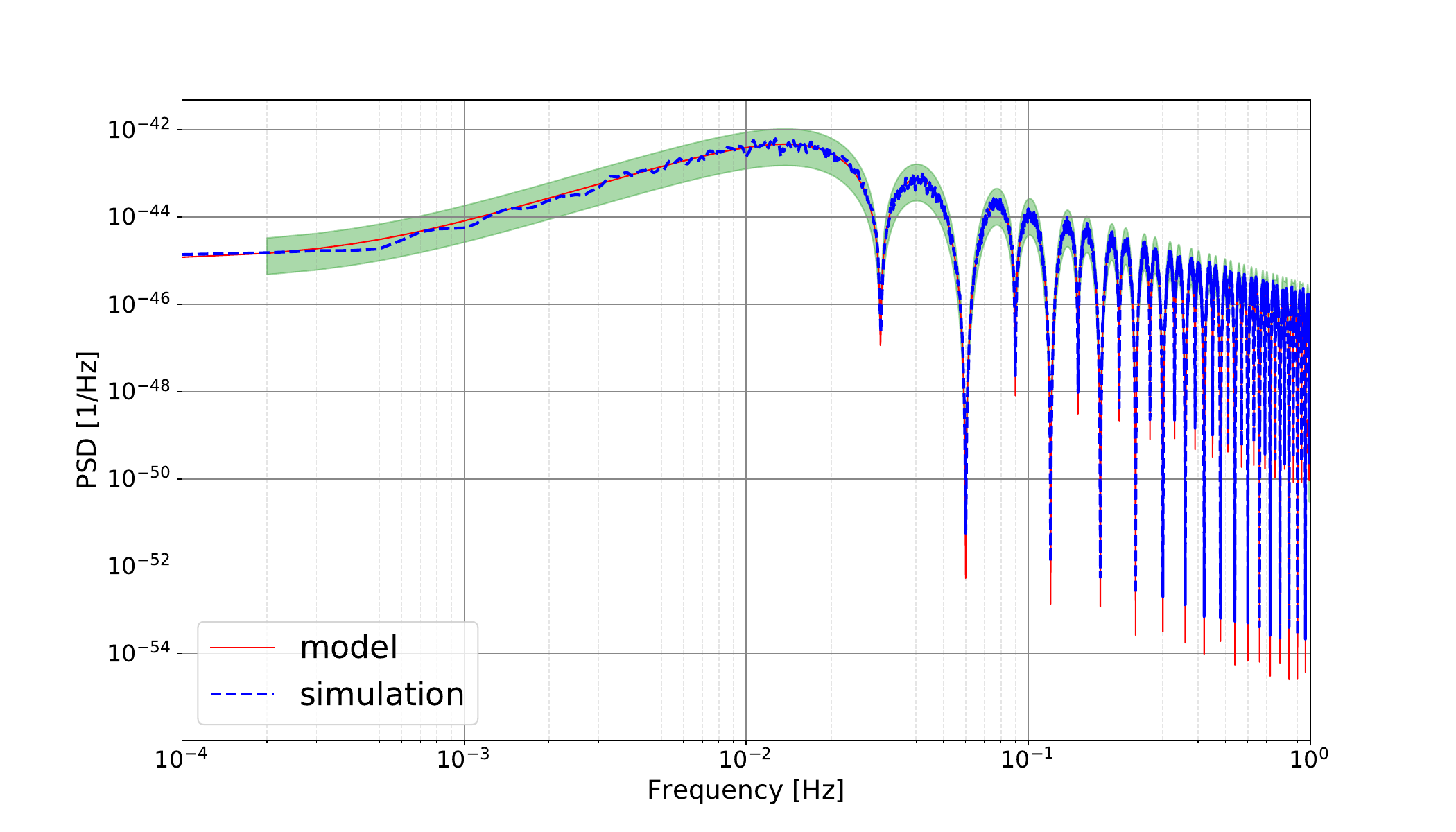}
    \caption{Correlated test-mass acceleration noise cross-comparison. The simulated data (red line) at 99.73\% confidence interval (green area) are in great agreement with the analytical formulation (blue dashed line).}
    \label{fig:validation_psd_accel_noise_corr}
    \end{center}
    \end{figure}
   
    \begin{figure}[h]
    \begin{center}
    \includegraphics[width=0.5\textwidth]{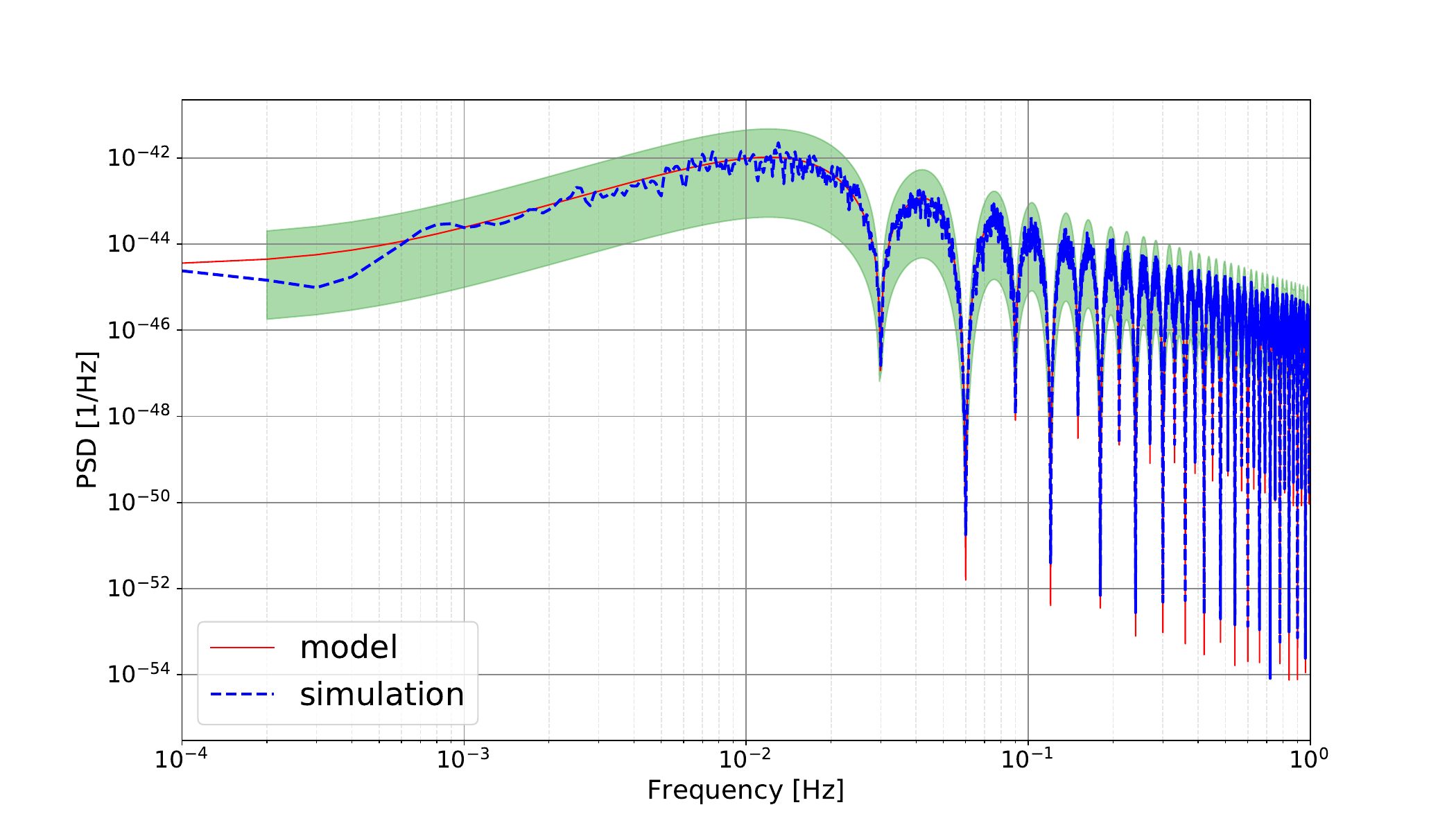}
    \caption{Anticorrelated test-mass acceleration noise cross-comparison. The simulated data (red line) at 99.73\% confidence interval (green area) are in great agreement with the analytical formulation (blue dashed line).}
    \label{fig:validation_psd_accel_noise_anti_corr}
    \end{center}
    \end{figure}

    \begin{figure}[h]
    \begin{center}
    \includegraphics[width=0.5\textwidth]{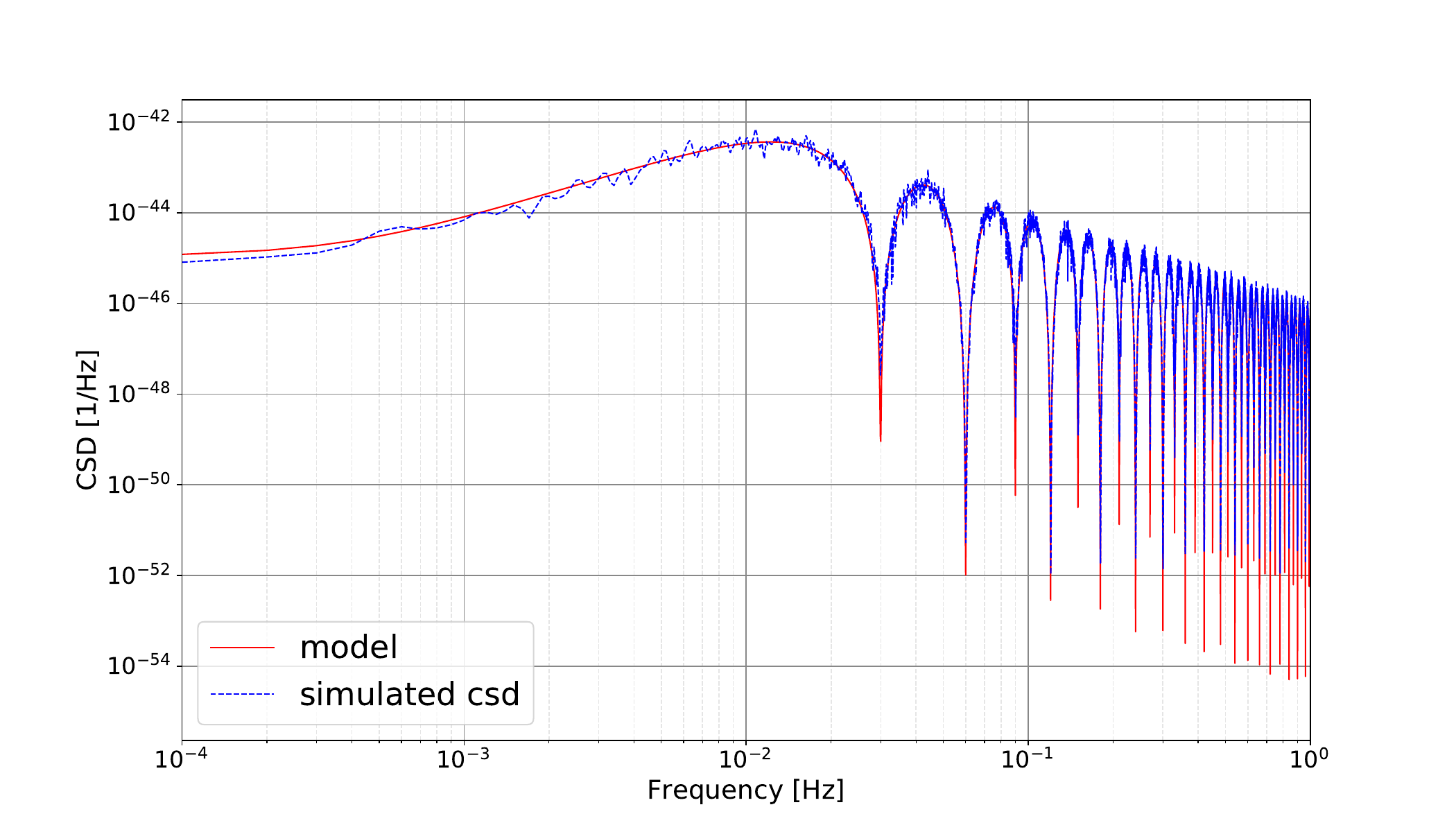}
    \caption{CSD uncorrelated TM acceleration noise. The red line is the simulated data and the blue dashed line is the analytical formulation.}
    \label{fig:validation_csd_accel_noise}
    \end{center}
    \end{figure}

 \subsection{About the propagation of suppressed noises}
 
 Although this article focuses on unsuppressed noises, for the sake of completeness, we will summarize the status of transfer functions for the suppressed noises, i.e., noises suppressed by \ac{TDI}, as well as the additional noises induced by this suppression.

Laser frequency noise has to be suppressed by several order of magnitude by \ac{TDI}, in order to be below the required noise level~\cite{SciRD,MRD,Babak:2021mhe} defined by the unsuppressed noises (acceleration, readout and optical path). 
It has been the main focus of \ac{TDI} noise reduction studies during many years, one of the most recent studies on the topic being~\cite{Bayle:2018hnm}. 
Because of the high level of reduction required, the residual level is sensitive to all limiting effects from the application of \ac{TDI}: flexing-filtering (noncommutation between antialiasing filters and delays)~\cite{Bayle:2018hnm}, ranging bias, stochastic ranging (imprecision in the knowledge of delays), interpolation, aliasing and fundamental armlength mismatch (limitation due to the flexing with \ac{TDI} 2.0). There are ongoing active studies on all these effects and preliminary transfer functions are already available enabling to establish the expected level of the residual laser noise.
Moreover, the residual laser noise depends on the laser locking configuration. Only preliminary checks based on simulation have been done and preliminary models have been developed~\cite{PhDHartwig}, and more detailed studies are necessary.

In principle, most effects leading to residual laser noise will also cause residuals in other noise sources which are perfectly canceled in an idealized situation. However, since these other suppressed noises are several orders of magnitude smaller than laser noise, their residuals can usually be neglected.

Clock noise is also reduced by \ac{TDI}. While its initial level is lower than that of laser noise, it is still a few orders of magnitude higher than the required noise level. 
In order to suppress clock noise, the laser beams carry sideband modulation with a clock-derived signal, creating so-called clock-sidebands. Interferometric measurements of these sidebands are then used in the \ac{TDI} algorithm to reduce clock noise~\cite{Hartwig:2020tdu}.

\ac{S/C} jitter noises $\va{\Delta}_{ij}$ are in theory perfectly canceled by \ac{TDI} when forming the $\xi_{ij}$ (see~\eqref{eq:tdi_xi12} and~\eqref{eq:tdi_xi13}). In reality, this cancellation will not be perfect and some residual noise is expected.

Finally, since the application of \ac{TDI} is a numerical procedure, some numerical limitations are expected.

The estimated residuals of all suppressed noises are currently below the required level, but some contributions are not negligible and need to be carefully studied.
The laser locking will impact some of these suppressed noises and is the topic of further studies currently underway.

\section{Conclusion}
\label{sec:conclusion}

The modeling of the noises and their propagation from the measurements to the \ac{TDI} variables are crucial for the LISA mission. Indeed, the \ac{TDI} algorithm will reduce some noise sources while leaving others largely untouched. The impact of correlations between links can either improve or deteriorate the performance of the mission at the \ac{TDI} level. We have seen this in the particular case of test mass acceleration noise, but it is also true for tilt-to-length~\cite{Paczkowski:2022nrt} or thermo-mechanical noises.  In addition, many noises related to the application of the algorithm itself, such as interpolation, clock noise residual or sideband modulation noise~\cite{Hartwig:2020tdu} can only be expressed at TDI level. Whether it is to establish the noise budget of the mission or to improve our understanding and knowledge of the noise for the needs of data analysis, the use of these TDI models is necessary.

The \ac{TDI} variables are the main data used to extract \ac{GW} signals. Therefore, it is important to have a good modeling of the noise \ac{PSD} and \ac{CSD} for the various \ac{TDI} variables in order to search for \ac{GW} sources, estimate their parameters and distinguish them from the instrument noises. This last point is particularly important for the search for stochastic gravitational wave backgrounds which can easily be confused with noise. 

In this article, we revisit a method to compute analytically the PSDs and the CSDs of unsuppressed noises at TDI level, as well as justify the approximations to simplify the result. We indeed derive the TDI transfer functions for most of noises in the update model for the LISA interferometric measurements in the more realistic configuration. We recover the transfer functions of the standard LISA unsuppressed noises \cite{SciRD, Estabrook:2000ef, Tinto:2014lxa} in the ideal case, i.e., under the assumption of equal armlengths and identical statistical properties of same type noises in different MOSAs. It also turns out that the optical and readout noises in the test-mass interferometers have different transfer function than the ones in reference and intersatellite interferometers. 

In addition, some standard cases of correlation have been studied. Accordingly, the spectral density of correlated noises could either improve or degrade the LISA noise budget, depending whether they are anticorrelated or fully correlated. Further analyses to identify correlation scenario preferable in reality are required.

The analytical expressions are provided in Tables~\ref{tab:summary_unsup_table_XYZ} and~\ref{tab:summary_unsup_table_AET} for the TDI variables $X, Y, Z, A, E$, and $T$. 
The analytical transfer functions of $X, Y, Z$ have been validated against simulations in different configurations.

The same method can be applied to any unsuppressed noises and to any \ac{TDI} variables.

The transfer functions for the unsuppressed noises with laser locking are the same as the ones without laser locking. It is not necessarily the case for suppressed noises, but we leave this for future works.
Actually the propagation of suppressed noises is usually more complicated. Several studies are underway and should soon result in publications.

\appendix

\section{Estimation of power spectral densities}
\label{app:EstimationPSD}
In the following we describe the procedure of estimating the power spectral density for a stochastic time series $x(t)$ of finite length $T$. We use the Scipy implementation of the so-called ``Welch's method''. It is summarized in the following steps. First, the data is divided into $M$ segments of length $L$ and a window function $w(t)$ is applied. Then, for each segment the Fourier transform is calculated which form independent estimates of the power spectral density as defined in \eqref{eq:psd_estimate}. Finally, the average [see \eqref{eq:psd_average}] over the $M$ segments is taken to reduce the variance.
\begin{align}
    \hat S^{(m)}(f_k) &= \frac{|\tilde x_w^{(m)}(f_k)|^2}{L} \label{eq:psd_estimate}\\
    \bar S(f_k) &= \frac{1}{M} \sum_{m=0}^{M-1}\hat S^{(m)}(f_k) \label{eq:psd_average}
\end{align}
This procedure yields estimates of $\bar S(f_k)$ at frequencies $f_k = \Delta f k$ with $k$ running from zero to $K=L f_s$. The spectral resolution is given by $\Delta f = \frac{1}{L}$. In theory one could choose to average over many segments to yield a very precise estimate of the PSD. However, in reality we are faced with limited amount of data and have to trade off between low variance and high spectral resolution.

In our studies we aim to validate the analytical \ac{PSD} models with simulated data. To check whether the  \ac{PSD} estimates $\bar S(f_k)$ are consistent with the model (null hypothesis) we conduct an hypothesis test. We define the confidence level $\gamma$ that represents the probability that all \ac{PSD} estimates are inside a given confidence interval.
\begin{equation}
    \gamma = \prod_{k}^{K-1} \mathrm{P}\left(\bar S_-(f_k) \le \bar S(f_k) \le \bar S_+(f_k)\right)
\end{equation}
We reject the null hypothesis if a single estimate $\bar S(f_k)$ resides outside the confidence interval.

The confidence intervals $[S_-(f_k), S_+(f_k)]$ can be derived from the statistics of the  \ac{PSD} estimates $\bar S(f_k)$. It is easy to show that $\bar S(f_k)$ has an expectation value of
\begin{equation}
    \mathrm{E}\{\bar S(f_k)\} = \frac{(|\tilde w|^2 * S)(f_k)}{L}
\end{equation}
Moreover, it has been demonstrated in \cite{jenkinsSpectralAnalysisIts1968a} that $\frac{\nu \bar S(f_k)}{\mathrm{E}\{\bar S(f_k)\}}$ is $\chi_\nu^2$ distributed with $\nu = 2 M$ degrees of freedom. By attributing ``equal confidence'' to each of the $K$ frequency bins we can write
\begin{equation}
    \mathrm{P}\left(\bar S_-(f_k) \le \bar S(f_k) \le \bar S_+(f_k)\right) = \gamma^\frac{1}{K} = 1 - \alpha
\label{eq:app_probConfInterv}
\end{equation}
where $\alpha$ is the probability that the estimate resides outside the confidence interval. The limits $\bar S_-(f_k)$ and $\bar S_+(f_k)$ are constructed symmetrically such that 
\begin{equation}
    \mathrm{P}\left(\bar S(f_k) < \bar S_-(f_k)\right) = \mathrm{P}\left(\bar S(f_k) > \bar S_+(f_k)\right) = \frac{\alpha}{2}
\end{equation}
They can be calculated by using the $\chi^2_\nu$ distributional property.

\begin{acronym}

\acro{AC}[AC]{Alternating Current}
\acro{AD}[AD]{Applicable Document}
\acro{ADC}[ADC]{Analog to Digital Converter}
\acro{AGN}[AGN]{Active Galactic Nuclei}

\acro{AIV}[AIV]{Assembly, Integration, Verification and validation}
\acro{AIVR}[AIVR]{Assembly, Integration, Verification and validation Requirements}
\acro{AIVT}[AIVT]{Assembly, Integration, Verification, and testing}
\acro{AIT}[AIT]{Assembly, Integration, and testing}
\acro{AK}[AK]{``Analytic Kludge''}
\acro{AKE}[AKE]{attitude absolute knowledge}
\acro{AMCVn}[AM~CVn]{class of cataclysmic variable stars}
\acro{AMR}[AMR]{Anisotropic Magnetoresistors}
\acro{AO}[AO]{Announcement of Opportunity}

\acro{AOCS}[AOCS]{Attitude and Orbit Control System}
\acro{AOM}[AOM]{Acousto-Optic Modulator}
\acro{ASD}[ASD]{Amplitude Spectral Density}
\acro{AST}[AST]{Autonomous Star Tracker}
\acro{AstroWG}[AstroWG]{Astrophysics Working Group}
\acro{ATA}[ATA]{\href{http://www.seti-inst.edu/ata/}{Allen Telescope Array}}
\acro{AU}[AU]{Astronomical Unit}

\acro{BAM}[BAM]{Beam Alignment Mechanism}
\acro{BAO}[BAO]{Baryonic Acoustic Oscillation}
\acro{BB}[BB]{Breadboard}
\acro{BBN}[BBN]{Big Bang nucleosynthesis}
\acro{BCRS}[BCRS]{Barycentric Celestial Reference System}
\acro{BEE}[BEE]{Back End Electronics}
\acro{BH}[BH]{Black Hole}
\acro{BHB}[BHB]{Black Hole Binary}
\acro{CAD}[CAD]{Computer Aided Design}
\acro{CAS}[CAS]{Constellation Acquisition Sensor}
\acro{CAC}[CAC]{Cost at Completion}
\acro{CATWP}[CATWP]{Catalogues Work Package}
\acro{CBE}[CBE]{Current Best Estimate}
\acro{CBOD}[CBOD]{Clamp band opening device}
\acro{CCD}[CCD]{Charge-coupled Device}
\acro{CCN}[CCN]{Contrat Change Notice}
\acro{CCU}[CCU]{Caging Control Unit}
\acro{CCPM}[CCPM]{Consortium Constellation Performance Model}
\acro{CDF}[CDF]{Concurrent Design Facility}
\acro{CDM}[CDM]{Cold dark Matter}
\acro{CDR}[CDR]{Critical Design Review}
\acro{CFRP}[CFRP]{Carbon Fibre Reinforced Plastic}
\acro{CFI}[CFI]{Consortium Furnished Item}
\acro{CM}[CM]{Caging Mechanism}
\acro{CMD}[CMD]{Charge Management Device}
\acro{CMM}[CMM]{Coordinate Measuring Machine}
\acro{CMS}[CMS]{Charge Management System}
\acro{CMNT}[CMNT]{Colloid Micro-Newton thruster}
\acro{CMB}[CMB]{Cosmic Microwave Background}
\acro{CNES}[CNES]{Centre National d’Etudes Spatiales}
\acro{COBE}[COBE]{\href{http://lambda.gsfc.nasa.gov/product/cobe/}{COsmic Background Explorer}}
\acro{CoM}[CoM]{Centre of Mass}
\acro{COMBO}[COMBO]{\href{http://www.mpia.de/COMBO/}{Classifying Objects by Medium-Band Observations}}
\acro{COSMOS}[COSMOS]{\href{http://irsa.ipac.caltech.edu/Missions/cosmos.html}{Cosmic Evolution Survey}}
\acro{CosWG}[CosWG]{Cosmology Working Group}
\acro{COTS}[COTS]{Commercial off the Shelf}
\acro{COTS}[COTS]{Commercial off the Shelf}
\acro{CPI}[CPI]{Consortium Provided Item}
\acro{CSD}[CSD]{cross power spectral density}
\acro{CSGS}[CSGS]{Consortium Science Ground Segment}
\acro{CTE}[CTE]{Coefficient of Thermal Expansion}
\acro{CTP}[CTP]{Core Technology Program}
\acro{CVM}[CVM]{Caging and Venting Mechanism}
\acro{DA}[DA]{Data Analysis}
\acro{DAAP}[DAAP]{Deep Analysis Alert Pipeline}
\acro{DAFTWP}[DAFTWP]{Data Analysis Framework and Tools Work Packages}
\acro{D/A}[D/A]{digital-to-analogue converter}
\acro{DCC}[DCC]{Data Computing Center}
\acro{DCCs}[DCCs]{Data Computing Centers}
\acro{DCP}[DCP]{De-Commissioning Phase}
\acro{DDE}[DDE]{Diagnostics Drive Electronics}
\acro{DDPC}[DDPC]{Distributed Data Processing Centre}
\acro{DEEP2}[DEEP2]{\href{http://deep.berkeley.edu/}{Deep Extragalactic Evolutionary Probe 2}}
\acro{DF}[DF]{drag-free}
\acro{DFACS}[DFACS]{drag--free attitude control system}
\acro{DHLC}[DHLC]{Data Handling and Laser Control}
\acro{DOF}[DOF]{degree of freedom}
\acro{DMU}[DMU]{Data Management Unit}
\acro{DMU}[DMS]{Document Management System}
\acro{DP}[DP]{diagnostic package}
\acro{DPC}[DPC]{Data Processing Centre}
\acro{DPEWG}[DPEWG]{Detection and Parameter Estimation Work Packages}
\acro{DPLL}[DPLL]{digital phase locked loop}
\acro{DRS}[DRS]{disturbance reduction system}
\acro{DSC}[Daughter-S/C]{``Daughter'' spacecraft}
\acro{DS}[DS]{Diagnostics Subsystem}
\acro{DSN}[DSN]{Deep Space Network}
\acro{DTM}[DTM]{deterministic transfer manoeuvre}
\acro{DWS}[DWS]{differential wavefront sensing}

\acro{E2E}[E2E]{End-to-End}
\acro{EBB}[EBB]{Elegant Breadboard}
\acro{ECSS}[ECSS]{European Cooperation for Space Standardization}
\acro{EC}[EC]{Executive Committee}
\acro{EDU}[EDU]{Engineering Development Unit}
\acro{EELV}[EELV]{Evolved Expendable Launch Vehicle}
\acro{EGAPS}[EGAPS]{European Galactic Plane Surveys}
\acro{EGSE}[EGSE]{Electrical Ground Support Equipment}
\acro{EH}[EH]{Electrode Housing}
\acro{EID}[EID]{Experiment Interface Document}
\acro{ELV}[ELV]{Expendable Launch Vehicle}
\acro{EMa}[EMa]{Electro-Magnetic}
\acro{EM}[EM]{Engineering Model}
\acro{EMC}[EMC]{Electro-Magnetic Contamination}
\acro{EMRI}[EMRI]{Extreme Mass-Ratio Inspiral}
\acro{EOL}[EOL]{End-Of-Life}
\acro{EoM}[EoM]{Equations of Motion}
\acro{EOM}[EOM]{Electro-Optical Modulator}
\acro{EPS}[EPS]{Extended Press-Schechter formalism}
\acro{ePMS}[ePMS]{extended Phase Measurement Subsystem}
\acro{EQM}[EQM]{Engineering and Qualification Model}
\acro{ESA}[ESA]{\href{https://www.esa.int/}{European Space Agency}}
\acro{ESAC}[ESAC]{European Space Astronomy Centre in Madrid, Spain}
\acro{ESOC}[ESOC]{European Space Operations Centre}
\acro{ESP}[ESP]{Extended Science Phase}
\acro{ESTEC}[ESTEC]{European Space Technology Centre}
\acro{ESTRACK}[ESTRACK]{European Space TRACKing}
\acro{ETU}[ETU]{Engineering Thermal Unit}
\acro{FAQ}[FAQ]{Frequently Asked Questions}
\acro{FBD}[FBD]{Functional Block Diagram}
\acro{FDIR}[FDIR]{Failure Detection, Isolation, and Recovery}
\acro{FDS}[FDS]{Frequency Distribution System}
\acro{FE}[FE]{finite-element (methods)}
\acro{FEE}[FEE]{front-end electronics}
\acro{FEEP}[FEEP]{field-emission electric propulsion}
\acro{FEESAU}[FEE SAU]{front-end electronics sensing and actuation unit}
\acro{FF-OGSE}[FF-OGSE]{Far-Field Optical Ground Support Equipment}
\acro{FITS}[FITS]{Flexible Image Transport System}
\acro{FIOS}[FIOS]{Fibre Injector Optical Subassembly}
\acro{FM}[FM]{Flight Model}
\acro{FMT}[FMT]{Formulation Management Team}
\acro{FOH}[FOH]{Fibre Optic Harness}
\acro{FPAG}[FPAG]{Fundamental Physics Advisory Group}
\acro{FPGA}[FPGA]{field-programmable gate array}
\acro{FPWG}[FPWG]{Fundamental Physics Working Group}
\acro{FR}[FR]{laser frequency  reference}
\acro{FRS}[FRS]{Frequency Reference System}
\acro{FS}[FS]{frequency  separated}
\acro{FSU}[FSU]{fibre switching unit}
\acro{FSUA}[FSUA]{fibre switching unit assembly}

\acro{GBs}[GBs]{Galactic Binaries}
\acro{GCR}[GCR]{Galactic Cosmic Ray}
\acro{GCRS}[GCRS]{Geocentric Celestial Reference System}
\acro{GRACE-FO}[GRACE-FO]{\href{https://gracefo.jpl.nasa.gov/~}{Gravity Recovery and Climate Explorer Follow On}}
\acro{GPRM}[GPRM]{Grabbing Positioning Release Mechanism}
\acro{GR}[GR]{General Theory of Relativity}
\acro{GRS}[GRS]{Gravitational Reference Sensor}
\acro{GRSH}[GRS]{Gravitational Reference Sensor Head}
\acro{GS}[GS]{Ground Station}
\acro{GSE}[GSE]{Ground Support Equipment}
\acro{GSFC}[GSFC]{Goddard Space Flight Center}
\acro{GTO}[GTO]{Geostationary Transfer Orbit}
\acro{GR740}[GR740]{\href{http://microelectronics.esa.int/gr740/index.html}{The ESA Next Generation Microprocessor (NGMP)}}
\acro{GW}[GW]{gravitational wave}

\acro{HDF}[HDF]{Hierarchical Data Format}
\acro{HDRM}[HDRM]{Hold Down and Release Mechanism}
\acro{HETO}[HETO]{Heliocentric Earth Trailing Orbit}
\acro{Hg}[Hg]{mercury}
\acro{HGA}[HGA]{high-gain antenna}
\acro{HR}[HR]{High Resolution}
\acro{HST}[HST]{\href{http://hubble.nasa.gov/~}{Hubble Space Telescope}}

\acro{IA}[IA]{Instrument Amplifier}
\acro{IAU}[IAU]{International Astronomical Union}
\acro{IAAS}[IAAS]{Infrastructure As A Service}
\acro{IBM}[IBM]{Internal Balance Mass}
\acro{ICC}[ICC]{Instrument Control Computer}
\acro{ICRF}[ICRF]{International Celestial Reference Frame}
\acro{ICRS}[ICRS]{International Celestial Reference System}
\acro{IDL}[IDL]{Interferometer Data Log}
\acro{IDS}[IDS]{Interferometric Detection System}
\acro{I/F}[I/F]{interface}
\acro{IFO}[IFO]{interferometer}
\acro{IFP}[IFP]{In-Field Pointing}
\acro{IGM}[IGM]{inter-galactic medium}
\acro{IMA}[IMA]{Integrated Modular Avionics}
\acro{IMBH}[IMBH]{Intermediate Mass Black Hole}
\acro{IMF}[IMF]{initial mass function}
\acro{IMR}[IMR]{Inspiral-Merger-Ringdown}
\acro{IMRI}[IMRI]{intermediate mass-ratio inspiral}
\acro{IMS}[IMS]{interferometric measurement system}
\acro{IN2P3}[IN2P3]{National Institute of Nuclear and Particle Physics}
\acro{INReP}[INReP]{Initial Noise Reduction Pipeline}
\acro{IOCR}[IOCR]{In-Orbit Commissioning Review}
\acro{IOT}[IOT]{Instrument Operations Team}
\acro{ISH}[ISH]{Inertial Sensor Head}
\acro{ISI}[ISI]{inter--satellite interferometer}
\acro{ISM}[ISM]{instrument sensitivity model}
\acro{ISO}[ISO]{International Organization for Standardization}
\acro{ISUK}[ISUK]{Inertial Sensor UV Kit}
\acro{IT}[IT]{Information Technology}
\acro{ITT}[ITT]{Invitation To Tender}

\acro{JILA}[JILA]{\href{http://jila.colorado.edu/}{Joint Institute for Laboratory Astrophysics}}
\acro{JPL}[JPL]{\href{http://www.jpl.nasa.gov/}{Jet Propulsion Laboratory}}
\acro{JWST}[JWST]{\href{http://www.jwst.nasa.gov/}{James Webb Space Telescope}}

\acro{KSC}[KSC]{\href{http://www.nasa.gov/centers/kennedy/home/index.html}{Kennedy Space Center}}

\acro{LA}[LA]{Laser Assembly}
\acro{LAGOS}[LAGOS]{Laser Antenna for Gravitational-radiation Observation in Space}
\acro{LCA}[LCA]{LISA Core Assembly}
\acro{LCM}[LCM]{NGO launch composite}
\acro{LDC}[LDC]{LISA Data Challenge}
\acro{LDC}[LDCWG]{LISA Data Challenge Working Group}
\acro{LDP}[LDP]{LISA Data Processing}
\acro{LDPG}[LDPG]{LISA Data Processing Group}
\acro{LED}[LED]{light-emitting diode}
\acro{LEM}[LEM]{Laser Electrical Module}
\acro{LEOP}[LEOP]{Launch and Early Operations Phase}
\acro{LGA}[LGA]{low-gain antenna}
\acro{LIG}[LIG]{LISA Instrument Group}
\acro{LIGO}[LIGO]{\href{http://www.ligo.caltech.edu/}{Laser Interferemeter Gravitational Wave Observatory}}
\acro{LISA}[LISA]{\href{https://www.lisamission.org/}{Laser Interferometer Space Antenna}}
\acro{LIST}[LIST]{\href{http://list.caltech.edu/}{LISA International Science Team}}
\acro{LLD}[LLD]{launch lock device}
\acro{LLP}[LLAP]{Low Latency Alert Pipeline}
\acro{LMC}[LMC]{Large Magellanic Cloud}
\acro{LMF}[LMF]{LISA mission formulation study}
\acro{LOA}[LoA]{Letter of Agreement}
\acro{LOM}[LOM]{Laser Optical Module}
\acro{LOS}[LOS]{line of sight}
\acro{LPF}[LPF]{\href{http://lisapathfinder.esa.int}{LISA Pathfinder}}
\acro{LPS}[LPS]{Laser Pre-stabilization System}
\acro{LTPDA}[LTPDA]{LISA Technology Package Data Analysis}
\acro{LH}[LH]{Laser Head}
\acro{LO}[LO]{Local Oscillator}
\acro{LRI}[LRI]{Laser Ranging Instrument (on GRACE-FO)}
\acro{LS}[LS]{laser system}
\acro{LSG}[LSG]{LISA Science Group}
\acro{LSGcore}[LSGcore]{LISA Science Group Core Team}
\acro{LSO}[LSO]{last stable orbit}
\acro{LSST}[LSST]{\href{http://www.lsst.org/lsst}{Large Synoptic Survey Telescope}}
\acro{LTP}[LTP]{LISA Technology Package}
\acro{LUT}[LUT]{Look-Up Table}
\acro{LVA}[LVA]{launch vehicle adaptor}
\acro{MAC}[MAC]{Mass Acceleration Curve}
\acro{MAG}[MAG]{Mission Analysis Guidelines}
\acro{MAXI}[MAXI]{\href{http://www.nasa.gov/mission_pages/station/research/experiments/MAXI.html}{Monitor of All-sky X-ray Image}}
\acro{MBH}[MBH]{Massive Black Hole}
\acro{MBHB}[MBHB]{Massive Black Hole Binary}
\acro{MCMC}[MCMC]{Markov-chain Monte Carlo}
\acro{MCR}[MCR]{Mission Consolidation Review}
\acro{MCS}[MCS]{Mission Control System}
\acro{MFR}[MFR]{Mission Formulation Review}
\acro{MCU}[MCU]{Mechanism Control Unit}
\acro{MEOP}[MEOP]{maximum expected operating pressure}
\acro{MGSE}[MGSE]{Mechanical Ground Support Equipment}
\acro{MIDA}[MIDA]{Mean Initial Displacement Angle}
\acro{MICD}[MICD]{Mechanical Interface Control Document}
\acro{MIRD}[MIRD]{Mission Implementation Requirement Document}
\acro{MIS}[MIS]{Mission}
\acro{MLA}[MLA]{Multi-lateral agreement}
\acro{MLB}[MLB]{motorised light band}
\acro{MLDC}[MLDC]{\href{http://astrogravs.nasa.gov/docs/mldc/}{Mock LISA Data Challenge}}
\acro{MLI}[MLI]{multi layer insulation}
\acro{MMAWP}[MMAWP]{Multi-Messenger Astrophysics Work Package}
\acro{MMH}[MMH]{monomethyl hydrazine}
\acro{MO}[MO]{Maser Oscillator}
\acro{MOAD}[MOAD]{Mission Operations Assumptions Document}
\acro{MOC}[MOC]{Mission Operation Centre}
\acro{MOFPA}[MOFPA]{Master Oscillator Fibre Power Amplifier}
\acro{MOPA}[MOPA]{Master Oscillator Power Amplifier}
\acro{MON-3}[MON-3]{mixed oxides of nitrogen with 3\% nitric oxide}
\acro{MOSA}[MOSA]{movable optical subassembly}
\acro{MOSAs}[MOSAs]{movable optical subassemblies}
\acro{MOU}[MoU]{Memorandum of Understanding}
\acro{MPR}[MPR]{Measured Pseudo-Range}
\acro{MRD}[LISA-MRD-001]{Mission Requirement Document}
\acro{MSC}[Mother-S/C]{``Mother'' spacecraft}
\acro{MSS}[MSS]{MOSA Selection Review}
\acro{MSS}[MSS]{MOSA Support Structure}

\acro{NASA}[NASA]{\href{http://www.nasa.gov}{National Aeronautic and Space Administration}}
\acro{NECP}[NECP]{Near-Earth Commissioning Phase}
\acro{NGMP}[NGMP]{Next Generation Micro Processor}
\acro{NGRM}[NGRM]{Next Generation Radiation Monitor}
\acro{NGO}[NGO]{New Gravitational wave Observatory}
\acro{NPMB}[NPMB]{National Program Managers Board}
\acro{NPRO}[NPRO]{Non-Planar Ring Oscillator}
\acro{NR}[NR]{Numerical Relativity}
\acro{NSP}[NSP]{Nominal Science Phase}
\acro{NTC}[NTC]{Negative Temperature Coefficient}
\acro{OAS}[OAS]{optical assembly subsystem}
\acro{OATM}[OATM]{Optical Assembly Tracking Mechanism}
\acro{OAM}[OAM]{optical assembly mechanics}
\acro{OB}[OB]{optical bench}
\acro{OBA}[OBA]{Optical Bench Assembly}
\acro{OBC}[OBC]{On-Board Computer}
\acro{OBDH}[OBDH]{On Board Data Handling}
\acro{OGS}[OGS]{Operational Ground Segment}
\acro{OGSE}[OGSE]{Optical Ground Support Equipment}
\acro{OM}[OM]{Optical Model}
\acro{OMS}[OMS]{Optical Metrology System}
\acro{OP}[OP]{optical path}
\acro{OPS}[OPS]{Operations}
\acro{ORO}[ORO]{optical read-out}
\acro{OT}[OT]{optical truss}

\acro{PA}[PA]{Power Amplifier}
\acro{PA}[PA]{Product Assurance}
\acro{PAA}[PAA]{Point-Ahead Angle}
\acro{PAAM}[PAAM]{Point-Ahead Angle Mechanism}
\acro{Pan-Starrs}[Pan-Starrs]{\href{http://pan-starrs.ifa.hawaii.edu/public/}{the Panoramic Survey Telescope \& Rapid Response System}}
\acro{PARD}[PARD]{Product Assurance Requirement Document}
\acro{PCU}[PCU]{Power Conditioning Unit}
\acro{PCDU}[PCDU]{Power Control and Distribution Unit}
\acro{PCP}[PCP]{Payload Commanding and Processing}
\acro{PCP-GSE}[PCP-GSE]{Payload Commanding and Processing Ground Support Equipment}
\acro{PCS}[PCS]{Payload Control Subsystem}
\acro{PDD}[PDD]{Payload Description Document}
\acro{PDF}[PDF]{Probability Density Function}
\acro{PDH}[PDH]{Pound Drever Hall}
\acro{PD}[PD]{Photo Diode}
\acro{PDR}[PDR]{Preliminary Design Review}
\acro{PDS}[PDS]{Photo Detector System}
\acro{PLAU}[PLAU]{Pre-Launch Phase}
\acro{PLS}[PLS]{Power Law Sensitivity}
\acro{PLL}[PLL]{Phase-Locked Loop}
\acro{P/L}[P/L]{Payload}
\acro{PL}[PL]{Payload}
\acro{P/M}[P/M]{propulsion module}
\acro{PM}[PM]{Progress Meeting}
\acro{PMF}[PMF]{Polarization-Maintaining Fibre}
\acro{PMFDE}[PMFDE]{Phase Meter Frequency Distribution Electronics}
\acro{PMFEE}[PMFEE]{Phase Meter Front-End Electronics}
\acro{PMDSP}[PMDSP]{Phase Meter Digital Signal Processor}
\acro{PMON}[PMON]{Power MONitor}
\acro{PMS}[PMS]{Phase Measurement Subsystem}
\acro{PN}[PN]{Post-Newtonian}
\acro{PPR}[PPR]{Proper Pseudo-Range}
\acro{PRR}[PRR]{Preliminary Requirements Review}
\acro{PRN}[PRN]{pseudo-random noise}
\acro{PRDS}[PRDS]{Phase Reference Distribution System}
\acro{PRDS-OGSE}[PRDS-OGSE]{Phase Reference Distribution System - Optical Ground Support Equipment}
\acro{PRT}[PRT]{Platinum Resistance Thermometers}
\acro{PSD}[PSD]{power spectral density}
\acro{PSF}[PSF]{point-spread function}
\acro{PTF}[PTF]{\href{http://www.astro.caltech.edu/ptf/}{Palomar Transient Factory}}

\acro{QA}[QA]{Quality Assurance}
\acro{QM}[QM]{Qualification Model}
\acro{QNM}[QNM]{Quasi-Normal Mode}
\acro{QPD}[QPD]{Quadrant photodetector}
\acro{QPR}[QPR]{Quadrant Photo-Receiver }
\acro{QSO}[QSO]{Quasi-stellar object}

\acro{RAAN}[RAAN]{Right Ascension of the Ascending Node}
\acro{RAM}[RAM]{Random-Access Memory}
\acro{RATS}[RATS]{Rapid Time Survey}
\acro{REF}[REF]{reference}
\acro{RF}[RF]{radio frequency}
\acro{RFI}[RFI]{reference interferometer}
\acro{RIN}[RIN]{Relative Intensity Noise}
\acro{RIT}[RIT]{Radio-frequency Ion Thruster}
\acro{RM}[RM]{Radiation Monitor}
\acro{RMS}[RMS]{Root Mean Square}
\acro{RS}[RS]{Requirement Specification}
\acro{RTOS}[RTOS]{Real Time Operating System}
\acro{RXTE}[RXTE]{\href{http://heasarc.gsfc.nasa.gov/docs/xte/xtegof.html}{Rossi X-Ray Timing Explorer}}
\acro{RX}[RX]{received signal}
\acro{S/C}[S/C]{spacecraft}
\acro{SC}[SC]{Spacecraft}
\acro{S/S}[S/S]{SubSystem}
\acro{S/C-P/M}[S/C-P/M]{spacecraft/propulsion-module}
\acro{SAVOIR}[SAVOIR]{\href{http://savoir.estec.esa.int/}{Space AVionics Open Interface aRchitecture}}
\acro{SAU}[SAU]{sensing and actuation unit}
\acro{SBBHs}[SBBHs]{Stellar Mass Black Hole Binaries}
\acro{SBCC}[SBCC]{Single Board Computer Core}
\acro{SCCP}[SCCP]{Science Commissioning and Calibration Phase}
\acro{SCET}[SCET]{Spacecraft Elapsed Time}
\acro{SCI}[SCI]{Science}
\acro{SciRD}[SciRD]{\href{https://www.cosmos.esa.int/documents/678316/1700384/SciRD.pdf/25831f6b-3c01-e215-5916-4ac6e4b306fb?t=1526479841000}{Science Requirement Document}}
\acro{SCOE}[SCOE]{Special Check Out Equipment}
\acro{SDP}[SDP]{system data pool}
\acro{SDSS}[SDSS]{\href{http://www.sdss.org/}{Sloan Digital Sky Survey}}
\acro{SEE}[SEE]{Single Event Error}
\acro{SEP}[SEP]{Solar Energetic Particle}
\acro{SEPD}[SEPD]{single-element photo diode}
\acro{SGS}[SGS]{Science Ground Segment}
\acro{SGWB}[SGWB]{Stochastic Gravitational Wave Background}
\acro{SIM}[SIM]{Space Interferometry Mission}
\acro{SIRD}[SIRD]{Science Implementation Requirement Document}
\acro{SIP}[SIP]{Science Implementation Plan}
\acro{SIPs}[SIPs]{Science Implementation Plans}
\acro{SIRD}[SIRD]{Science Implementation Requirements Document}
\acro{SISO}[SISO]{Single Input / Single Output}
\acro{SIWP}[SIWP]{Science Interpretation  Work Package}
\acro{SL}[SL]{Scatter Light}
\acro{SMF}[SMF]{Single Mode Fiber}
\acro{SQUID}[SQUID]{Superconducting Quantum Interference Device}
\acro{SMBH}[SMBH]{super-massive black hole}
\acro{SMC}[SMC]{Small Magellanic Cloud}
\acro{SMP}[SMP]{Science Management Plan}
\acro{SNR}[SNR]{Signal-to-Noise Ratio}
\acro{SOBH}[SOBHB]{Stellar Origin Black Hole Binary}
\acro{SOCD}[SOCD]{Science Operations Concept Document}
\acro{SOAD}[SOAD]{Science Operations Assumptions Document}
\acro{SOC}[SOC]{Science Operation Centre}
\acro{SOSL}[SOSL]{Science Operation Study Lead}
\acro{SOVT}[SOVT]{Science Operations Verifications Tests}
\acro{SPA}[SPA]{Stationary Phase Approximation}
\acro{SPC}[SPC]{Science Programme Committee}
\acro{SRR}[SRD]{System Requirement Document}
\acro{SRR}[SRR]{System Requirement Review}
\acro{SRP}[SRP]{Solar Radiation Pressure}
\acro{SRS}[SRS]{Spacecraft Reference System}
\acro{SSB}[SSB]{Solar System Barycenter}
\acro{SST}[SST]{Science Study Team}
\acro{STM}[STM]{Structural and Thermal Model}
\acro{STOP}[STOP]{Structural and Thermal Optical Performance}
\acro{STR}[STR]{Coarse Star Tracker}
\acro{SW}[SW]{Software}
\acro{SWT}[SWT]{Science Working Team}

\acro{TBC}[TBC]{To Be Confirmed}
\acro{TBD}[TBD]{To Be Determined}
\acro{TBW}[TBD]{To Be Written}
\acro{TC}[TC]{Telecommand}
\acro{TCB}[TCB]{Barycentric Coordinate Time}
\acro{TCG}[TCG]{Geocentric Coordinate Time}
\acro{TCM}[TCM]{trajectory correction manoeuvre}
\acro{TC/TM}[TC/TM]{telecommand/telemetry}
\acro{TCLS}[TCLS]{\href{http://www.tcls-arm-for-space.eu/} {Triple Core LockStep}}
\acro{TDI}[TDI]{Time delay interferometry}
\acro{TECC}[TECC]{Transient Event Coordination Committee}
\acro{TID}[TID]{Total Ionising Dose}
\acro{THE}[THE]{On-Board Clock Time}
\acro{TM}[TM]{test mass, \emph{often proof mass}}
\acro{TMI}[TMI]{test--mass interferometer}
\acro{TM-OGSE}[TM-OGSE]{Test-Mass Optical Ground Equipment}
\acro{TNO}[TNO]{Nederlandse Organisatie voor Toegepast Natuurwetenschappelijk Onderzoek}
\acro{TNID}[TNID]{Total Non-Ionizing Dose}
\acro{TOBA}[TOBA]{Telescope and Optical Bench Assembly}
\acro{TOGA}[TOGA]{Telescope, Optical bench and Gravitational reference sensor Assembly}
\acro{TSP}[TSP]{Temporal and Spatial Partitioning}
\acro{TP}[TP]{Telescope Pointing}
\acro{TP}[TP]{Transfer Phase}
\acro{TPS}[TPS]{Spacecraft Proper Time}
\acro{TRA}[TRA]{Technology Readiness Assessment}
\acro{TRL}[TRL]{Technology Readiness Level}
\acro{TRP}[TRP]{Temperature Reference Points}
\acro{TS}[TS]{Telescope}
\acro{TTC}[TT\&C]{Telemetry, Tracking, and Command}
\acro{TTL}[TTL]{Tilt-To-Length}
\acro{TWTA}[TWTA]{traveling-wave tube amplifier}
\acro{TX}[TX]{transmit signal}
\acro{UCB}[UCB]{Ultra Compact Binary}
\acro{ULU}[ULU]{UV light unit}
\acro{US}[US]{United States (of America)}
\acro{USO}[USO]{Ultra-Stable Oscillator}
\acro{UTC}[UTC]{Universal Time Coordinated}
\acro{UV}[UV]{Ultra-Violet}
\acro{VAST}[VAST]{\href{http://www.physics.usyd.edu.au/sifa/vast/index.php}{Variables and Slow Transients, An ASKAP Survey for Variables and Slow Transients is a Survey Science Project for the Australian SKA Pathfinder}}
\acro{VB}[VB]{Verification Binary}
\acro{VC}[VC]{Vacuum Chamber}
\acro{VGB}[VGB]{Verification Galactic Binary}
\acro{VGBs}[VGBs]{Verification Galactic Binaries}
\acro{VMS}[VMS]{Very Massive Star}
\acro{WAVWG}[WavWG]{Waveform Working Group}
\acro{WD}[WD]{White Dwarf}
\acro{WG}[WG]{Working Group}
\acro{WMAP}[WMAP]{\href{http://lambda.gsfc.nasa.gov/product/map/current/}{Wilkison Microwave Anisotropy Probe}}
\acro{WP}[WP]{Work Package}
\acro{WPs}[WPs]{Work Packages}
\acro{WR}[WR]{Wide Range}

\acro{XML}[XML]{Extensible Markup Language}



\end{acronym}

\acknowledgments
The authors thank Gerhard Heinzel for the fruitful exchanges.
The authors also thank the Performance Working/Expert Group and the Simulation Working/Expert Group of the LISA Consortium. 
This work is supported by the Centre National d’Études Spatiales (CNES), the Centre National de la Recherche Scientifique (CNRS), the Universit\'e Paris Cit\'e (former Universi\'e Paris Diderot), the Institut de la Recherche sur les lois Fondamentales de l'Univers of the Commissariat \`a l'\'Energie Atomique et aux \'energies alternatives (CEA/IRFU) and the Observatoire de Paris.
It was also supported by the Programme National GRAM of CNRS/INSU with INP and IN2P3 co-funded by CNES.

\bibliographystyle{apsrev4-1}
\bibliography{refs}

\end{document}